\documentclass[final]{elsarticle}
\usepackage{amssymb}

\begin{document}
\def\bfs{\mathbf{S}}
\def\bfk{\mathbf{k}}
\def\bft{\mathbf{t}}
\def\bfe{\mathbf{e}}
\def\bfx{\mathbf{x}}
\def\tr{\mathrm{tr}}

\begin{frontmatter}

\title{Critical density of topological defects upon a continuous phase transition}

\author{A.O. Sorokin}
\ead{aosorokin@gmail.com}

\address{Petersburg Nuclear Physics Institute, NRC Kurchatov Institute, 188300 Orlova Roscha, Gatchina, Russia}

\begin{abstract}
Using extensive Monte Carlo simulations, we test the hypothesis that the density of corresponding topological defects has an universal value at the temperature of a continuous phase transition. We consider several simple two-dimensional models where domain walls, vortices, so-called $\mathbb{Z}_2$ vortices or their combinations are presented. These topological defects are relevant correspondingly to an Ising second-order phase transition, a Berezinskii-Kosterlitz Thouless transition and an explicit crossover. We compare results for square and triangular lattices as well as for the complicated situation when two types of defects are presented and two transitions occur separated in temperature. All considered cases demonstrate consentient results confirming the hypothesis.
\end{abstract}


\end{frontmatter}

\section{Introduction}

Topological defects play an important role in the theory of phase transition and critical phenomena. The most famous example is vortex tubes arising in type-II superconductors \cite{Abrikosov57}, superfluid $^4$He \cite{Onsager49,Feinman55}, XY ferromagnets \cite{Kohring86,Kohring87} and cosmological models \cite{Zurek85}. Nevertheless, the modern theory of critical phenomena, based primarily on the renormalization group (RG) approach, even though describes the critical behavior correctly but does not bring out the explicit contribution of topological defects. The important exception is a Berezinskii-Kosterlitz-Thouless (BKT) transition taking place in two-dimensional XY ferromagnets \cite{Berezinsky71,Berezinsky72,Kosterlitz73,Kosterlitz74}. In this case, one can distinguish the part of the partition function corresponding to a system of vortices interacting like Coulomb charges \cite{Berezinsky71,Villain75,Kadanoff77} and obtain RG-equations that correctly describe a BKT transition. Although some generalizations of a BKT transition are known (see, e.g., \cite{Banks77,Cardy80}), in the general case the direct contribution of topological defects to the critical behavior remains unclear.

The evidence that topological defects are essential for the critical behavior is given by systems with the same local geometry of an order parameter space $G/H$, but with a different topology of it. For example, in the three-dimensional $O(4)$ model describing classical ferromagnets with a four-component spin ($N=4$), a second-order transition occurs, whereas in the model with $G/H=SO(3)$ (describing, e.g., frustrated helimagnets \cite{Sorokin14}), a transition is of first order induced by fluctuations. The local geometry of the order parameter spaces in these models are the same $SO(4)/SO(3)\sim SO(3)$. It means that the models have the same spectrum of Goldstone modes, similar low-temperature behavior, and the same critical behavior close to two dimensions \cite{Hikami81}. However, the topologies of these spaces are different. The space $SO(4)/SO(3)$ is equivalent to a four-dimensional sphere $S^3$, but $SO(3)\approx S^3/\mathbb{Z}_2$ with the nontrivial fundamental group $\pi_1(SO(3))=\mathbb{Z}_2$. In the second case, there are topologically stable configurations in the spectrum, called $\mathbb{Z}_2$ vortices.

Meanwhile, topological defects are in abundance represented in magnets both in two and three dimensions: domain walls in the Ising model ($N=1$), vortices in XY magnets ($N=2$), $\mathbb{Z}_2$ vortices in magnets with a non-collinear spin ordering, skyrmions and even monopoles in a three-dimensional isotropic ($N=3$) ferromagnets. (Skyrmion-like topological defects are not considered in this paper.) Of course, the presence of topologically stable configurations does not determine a type of the critical behavior. One knows examples of systems with defects, where transition is of either first or second order, as well as a transition with a BKT-behavior (infinite order), or a phase transition is absent.

The common property of topological defects of any type is that they bring additional disorder into a system. Within a defect core (with exception of skyrmion-like configurations), a symmetry, broken in the ordered phase, is restored. Thus, an increase in the concentration of defects should lead to the destruction of an order. This is observed in type-II superconductors. There is such a value of defect density when an long-range (or quasi-long-range) order disappears. In the work \cite{Antunes01} it has been supposed that such a critical density of defects may have a universal value. This hypothesis, partially confirmed experimentally, is based on extrapolation of the Halperin's formula \cite{Halperin80,Liu92} for the density of Green function zeros to the critical region.

The idea of the Halperin's formula is based on following. Since in the defect core a symmetry is restored, a value of the order parameter vanishes $|\varphi(r)|=0$. Thus the core of a topological defect corresponds to coincident zeros of the two-point function $G(x)=\langle\varphi(0)\varphi(x)\rangle$. For an $O(N)$ symmetric model with a Gaussian field $\varphi$, the density of Green function zeros is \cite{Halperin80,Liu92}
\begin{equation}
    \rho=A\left|\frac{G''(0)}{G(0)}\right|^{N/2},
    \label{r=g}
\end{equation}
where $A$ is a coefficient, which can be calculated if the field $\varphi(x)$ is still Gaussian. The authors of \cite{Antunes01} have proposed to extrapolate this formula to the critical region, where the two-point function behaves as
\begin{equation}
    G(x)\sim\int d^dk\frac{e^{ikx}}{|k|^{2-\eta}},
\end{equation}
with $\eta$ is the anomalous dimension of the field $\varphi$. So
\begin{equation}
    \rho=A\left|\frac{\eta+d-2}{\eta+d}\right|^{N/2}.
    \label{r=a-eta}
\end{equation}
The coefficient of proportionality $A$ have been obtained \cite{Antunes01} using the infinite temperature limit $T\to\infty$ with $\eta=2$, where the density of defects can be calculated in the alternative way \cite{Vilenkin84}. In this case, the coefficient $A$ does not depend on a lattice type, the Fisher's index $\eta$ is universal, and hence the density of topological defects (\ref{r=a-eta}) at the critical point has a universal value.

There are a few arguments against this hypothesis. The main one that the field $\varphi$ describing fluctuations of the order parameter is not Gaussian in the vicinity of the critical point. Also, as we show below, additional difficulties arise when one has several types of topological defects or several subsequent phase transitions.

However, since the hypothesis is partially confirmed experimentally, in this paper we carry out its additional verification. Using extensive Monte Carlo simulations, we study several models with different number and types of topological defects, as well as with different numbers and types of (continuous) transitions: a second-order transition, a BKT transition and a crossover close to a second-order transition. To check the universality, we consider these models both on square and triangular lattices. Only the case of two dimensions are investigated.

We estimate the critical value of the density for domain walls, vortices and $\mathbb{Z}_2$ vortices using the Ising, $O(2)$ and $V_{3,2}$ Stiefel models correspondingly. In addition, we study models where two types of defects are presented and two sequential transition occur. Among these models, the Ising-XY has been investigated earlier, but the rest are investigated for the first time.

\section{Ising model}

The simplest case is the Ising model where topological defects are line-like domain walls. This case is very useful for the hypothesis confirmation due to two reasons. First, the Ising model can be easy reformulated in terms of domain walls. And second, we have the exact solution of the model \cite{Onsager44}.

The Hamiltonian of the Ising model is
\begin{equation}
  H=-J\sum_{ij} s_{i}s_{j},\quad s_{i}=\pm 1,
\end{equation}
where the sum $ij$ runs over neighboring sites of a lattice. The domain wall density is defined as
\begin{equation}
    \tilde\rho_\mathrm{dw}=\frac{1}{2pL^2}\sum_{ij}(1-s_i s_j),
    \quad \rho_\mathrm{dw}=\langle\tilde\rho_\mathrm{dw}\rangle,
    \label{ro-dw-ising}
\end{equation}
where $p=2,\,3$ for a square and triangle lattice correspondingly, $L^2$ is a lattice volume. So the domain wall density relates to the eternal energy as
\begin{equation}
    E=JpL^2\left(-1+2\rho_\mathrm{dw}\right).
\end{equation}
Lets use the exact solution of the Ising model. We set $J=1$ for simplicity. The free energy of the model on a square lattice is \cite{Houtappel50}
$$
 \ln Z=-\frac{F(\square)}{T L^2}=-\ln2 + \ln\left(1-\tanh^2K\right)-
$$
\begin{equation}
 -\frac1{8 \pi^2} \int\limits_0^{2\pi}\int\limits_0^{2\pi}dadb\,
   \ln\left(\left(1 + \tanh^2K\right)^2-2\tanh K \left(1 - \tanh^2K\right)(\cos a + \cos b)\right),
\end{equation}
where $K=\frac{1}{T}$, $Z=\sum\exp(-H/T)$. The free energy of the model on a triangle lattice is \cite{Houtappel50}
$$
 \ln Z=-\frac{F(\triangle)}{T L^2}=-\ln 2-
$$
\begin{equation}
 -\frac1{8 \pi^2} \int\limits_0^{2\pi}\int\limits_0^{2\pi} dadb\,
   \ln\left(\cosh^3 2K + \sinh^3 2K - \sinh 2K (\cos a + \cos b  + \cos(a + b))\right).
\end{equation}
Using numerical integration of the solutions, the formula
\begin{equation}
    E=\frac{T^2}{Z}\frac{\partial Z}{\partial T},
\end{equation}
and the values of the critical temperature \cite{Kramers41}
\begin{equation}
    T_c(\square)=\frac{2}{\ln(\sqrt{2}-1)}\approx 2.269185\ldots,\quad
    T_c({\triangle})=\frac{2}{\ln(\sqrt{3})}\approx 3.6409569\ldots,
\end{equation}
we obtain the critical value of the defect density
\begin{equation}
\rho_\mathrm{dw}(\square)=0.1464466\ldots,\quad
\rho_\mathrm{dw}(\triangle)=0.1666666\ldots.
\end{equation}
We immediately note that these values do not agree with naive use of the extrapolated formula (\ref{r=a-eta}) with $\eta=1/4$ and $A=1/\pi$: $\rho_\mathrm{dw}\approx0.106$. It also does not agree with results of the procedure \cite{Antunes01} with $\rho_\mathrm{dw}(T\to\infty)=0.5$: $\rho_\mathrm{dw}\approx0.2357$.

\begin{figure}[t]
\center
a)
\includegraphics[scale=0.35]{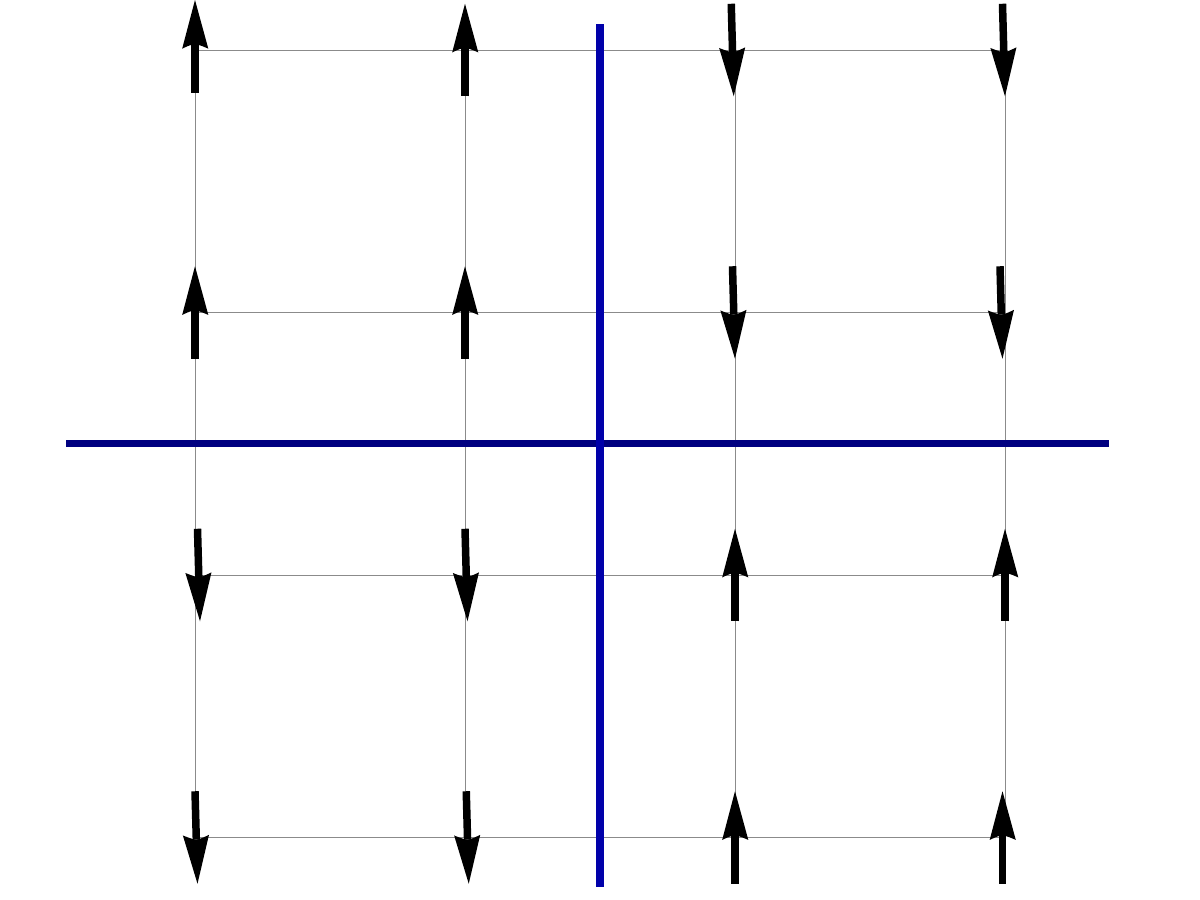}
b)
\includegraphics[scale=0.35]{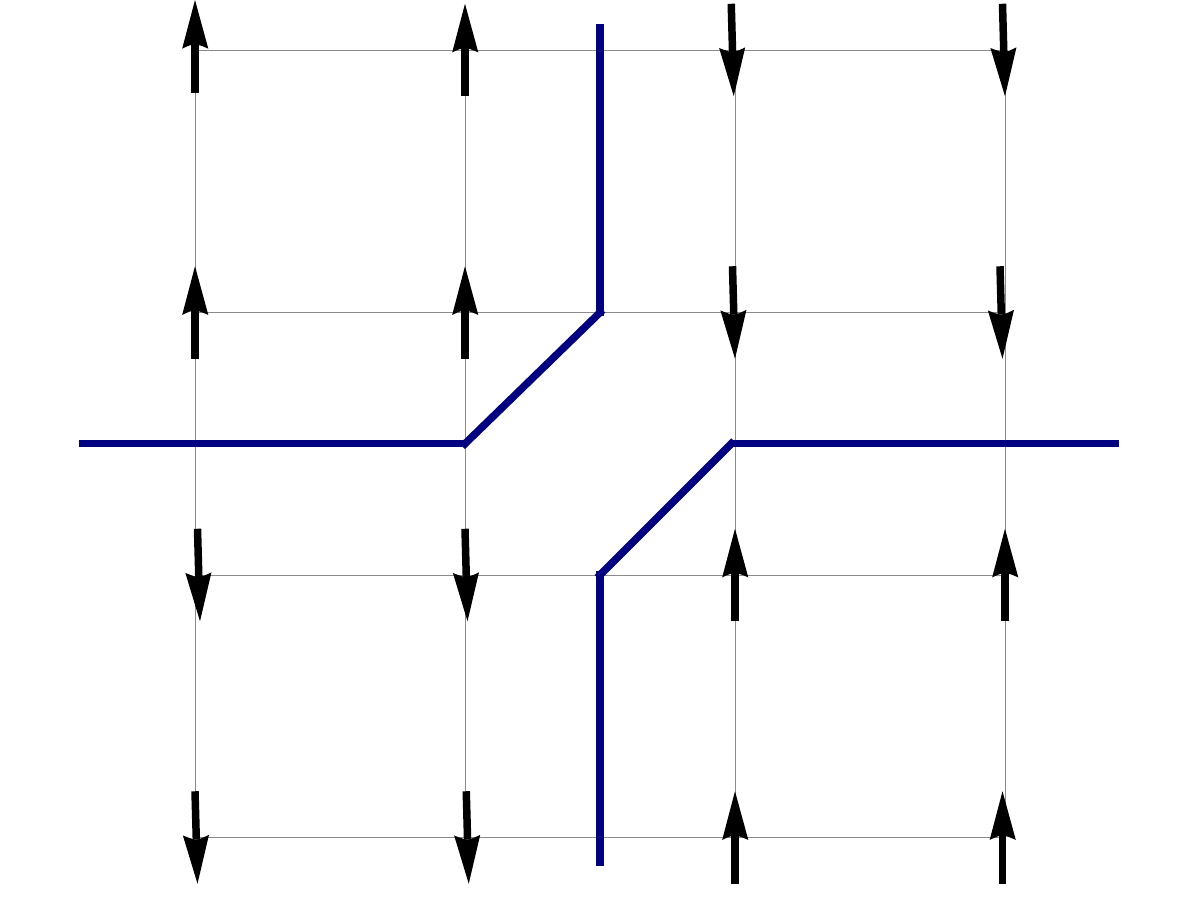}
\caption{\label{fig1} a) X-junction of domain walls, presented on a square lattice; b) its possible interpretation.}
\end{figure}%
To compare the values of the domain wall density for both lattices, it is necessary to take into account that length of the unit element of a wall on a triangle lattice is $\sqrt3/2$ times smaller than on a square lattice. (More precisely, a distance between sites of a dual (honeycomb) lattice is $\sqrt3/3$, but a number of wall elements is $3/2$ times larger.) So the density ratio is
\begin{equation}
    r_\mathrm{dw}=\frac{2}{\sqrt{3}}\frac{\rho_\mathrm{dw}(\square)}{\rho_\mathrm{dw}(\triangle)}=1.01461\ldots,
    \label{ratio-ising-old}
\end{equation}
which is close to 1, but not equal exactly.

In fact, our definition of the wall density (\ref{ro-dw-ising}) on a lattice is not fully correct. It does not lead to the unified continuous limit for different lattices when lattice constant vanishes. Our definition measures the total length of all walls and divides it by a number of unit cells. We remind that at the critical temperature $T=T_c$ and thermodynamical limit $L\to\infty$, there are two types of walls: close and infinite. Walls of the last type are boundaries of infinite clusters. The theory based on the Schramm-Loewner evolution gives some information about the critical properties of such walls (see \cite{Cardy05} for a review). Due to zero tension of walls at the critical point, walls have non-trivial (fractal) form with the Hausdorff dimension being not an integer, namely $D_f=11/8$. Of course, the fractal dimension of a wall can be determined in a lattice theory, the index $D_f$ shows how the length of the boundary of the largest cluster grows with increasing $L$: $l(L)\sim L^{D_f}$. On a triangular lattice, this procedure can be carried out and leads to the correct result, since the length of a wall is uniquely defined. But on a square lattice, a wall may be self-intersecting, so the definition of a cluster boundary is not unique. Such self-intersections (or X-junction) of walls (see fig. \ref{fig1}(a)) have no analogue on a triangular lattice and arise only in the continuous limit.

To make the procedure for determining the wall density more correct, it is required to propose the replacement of wall intersections by a certain configuration of disjoint walls. An X-junction of walls has double topological charge $\tilde\rho_\mathrm{dw}=2$ and adds two unit length to the total length. A non-intersection walls configuration adds a surely smaller value, so the ratio (\ref{ratio-ising-old}) becomes smaller too. For example, the configuration shown in fig. \ref{fig1}(b) adds the length $\tilde\rho=\sqrt2\approx1.4$. Also, to complete the procedure, one should know the density of X-junctions at the critical point, but this quantity cannot be obtain using the exact solutions.

\begin{figure}[t]
\center
a)
\includegraphics[scale=0.35]{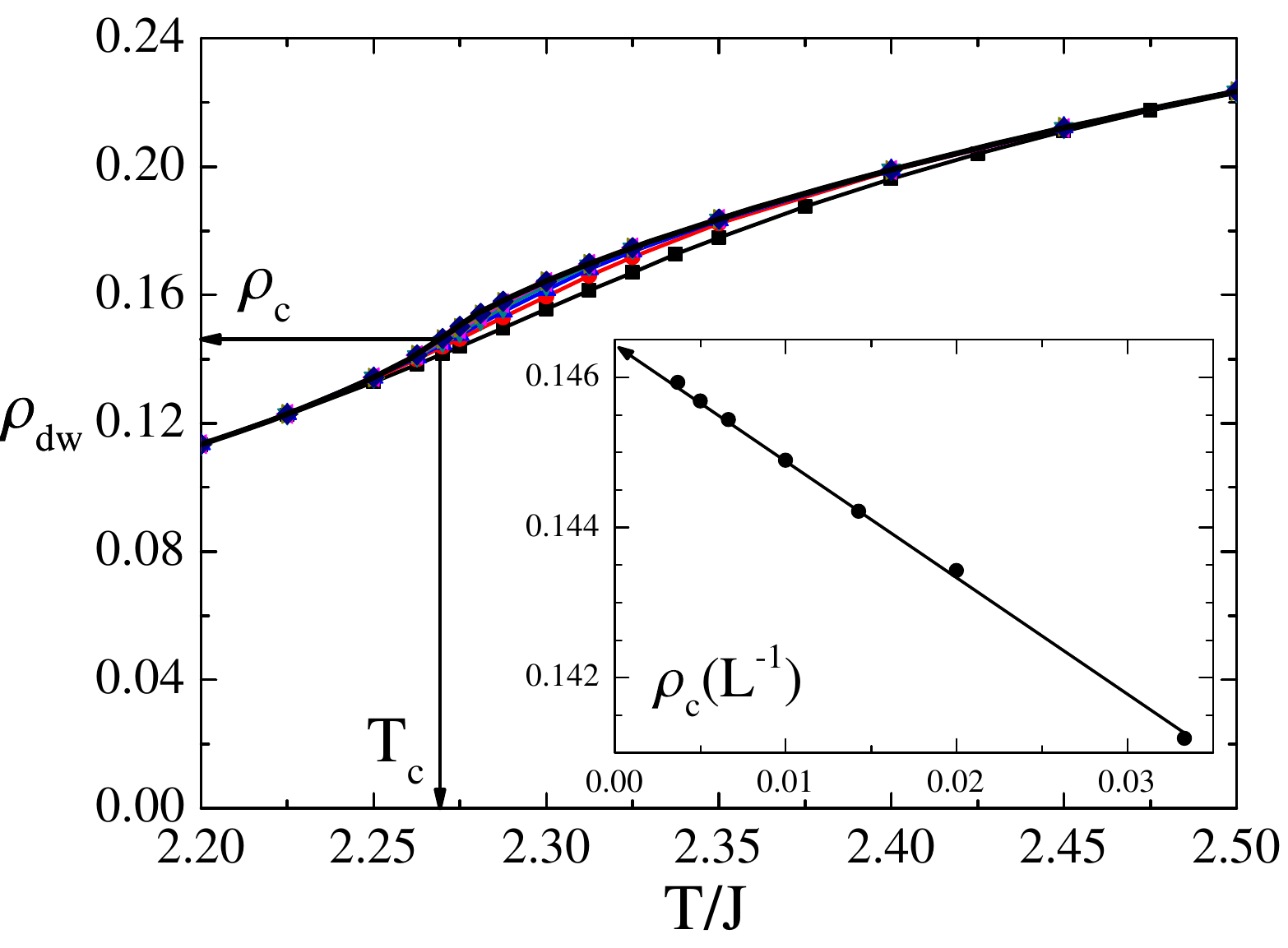}
b)
\includegraphics[scale=0.35]{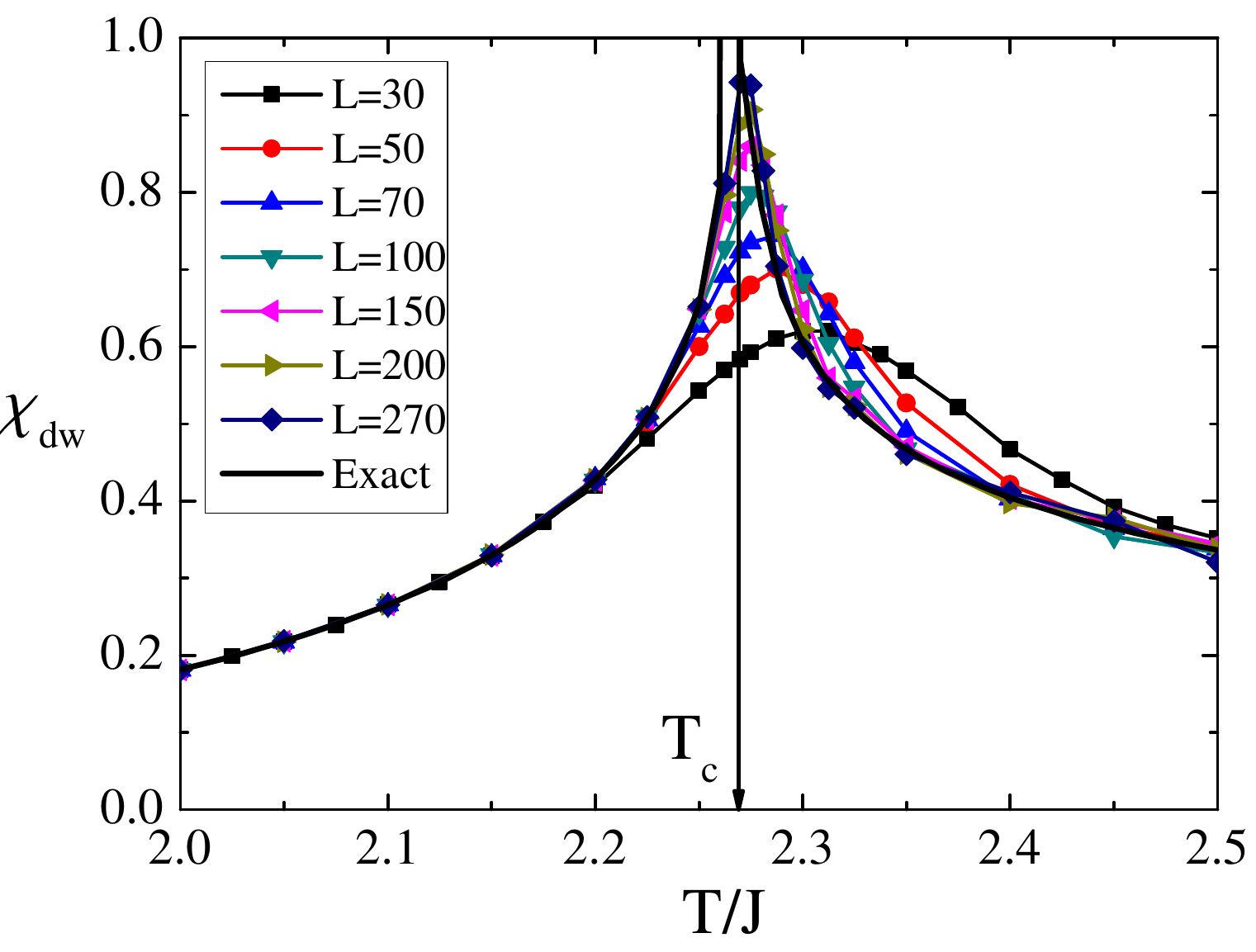}
\caption{\label{fig2} Domain wall density (a) and topological susceptibility (b) in the Ising model on a square lattice.}
\end{figure}%
To estimate the density of wall X-junctions numerically, we perform Monte Carlo simulations of the Ising model. We use the Wollf cluster algorithm \cite{Wollf89}, a periodic boundary conditions, square and triangular lattices with $L=30,$ 50, 70, 100, 150, 200 and 270. Thermalization is performed within $3\cdot10^5$ Monte Carlo steps per spin, and calculation of averages within $6\cdot10^6$ steps.

In addition to the density of domain walls (\ref{ro-dw-ising}) and their X-junctions
\begin{equation}
\tilde\rho_\mathrm{xj}(x)=\frac1{16}\prod_{\square_x}(1-s_i s_j),\quad \tilde\rho_\mathrm{xj}=\sum_x\tilde\rho_\mathrm{xj}(x),\quad \rho_\mathrm{xj}=\langle\tilde\rho_\mathrm{xj}\rangle,
\label{X-junction}
\end{equation}
where $x$ runs over square cells of a lattice, we also monitor the quantity called topological susceptibility
\begin{equation}
\chi_\mathrm{td}=L^2\left(\langle\rho_\mathrm{td}^2\rangle-\langle\rho_\mathrm{td}\rangle^2\right).
\label{chi}
\end{equation}
It indicates a relevance of defects to a considered phase transition. So, if it has a singularity, defects are significant at a critical point. In the Ising model, the topological density and susceptibility behave as (fig. \ref{fig2})
\begin{equation}
    \rho_\mathrm{dw}=\rho(L)+C_1 L^{-\beta_\mathrm{dw}},\quad
    \chi_\mathrm{dw}(L)=C_2+C_3 \ln L, \quad
    T=T_c,
\end{equation}
where $\beta_\mathrm{dw}=1$. Numerical estimation of critical densities is
\begin{equation}
\rho_\mathrm{dw}(\triangle)=0.16663(3),\quad
\rho_\mathrm{dw}(\square)=0.14644(2),\quad
\rho_\mathrm{xj}=0.00494(2).
\end{equation}
The corrected density ratio is
\begin{equation}
    r_\mathrm{dw}=\frac{2}{\sqrt{3}}\frac{\rho_\mathrm{dw}(\square)-q\rho_\mathrm{xj}}{\rho_\mathrm{dw}(\triangle)},
    \label{ratio-ising-new}
\end{equation}
where $q$ is relative difference between lengthes of walls in an X-junction and substitutive non-intersecting configuration. Since, $q=0$ for the X-junction (fig. \ref{fig1}(a)) with the maximal length, and $q=2-\sqrt2$ for the configuration (fig. \ref{fig1}(b)) with the minimal length, we see
\begin{equation}
0.99456\leq r_\mathrm{dw}\leq1.01461,
\end{equation}
that contains the case $r_\mathrm{dw}=1$.
\begin{figure}[t]
\center
a)
\includegraphics[scale=0.35]{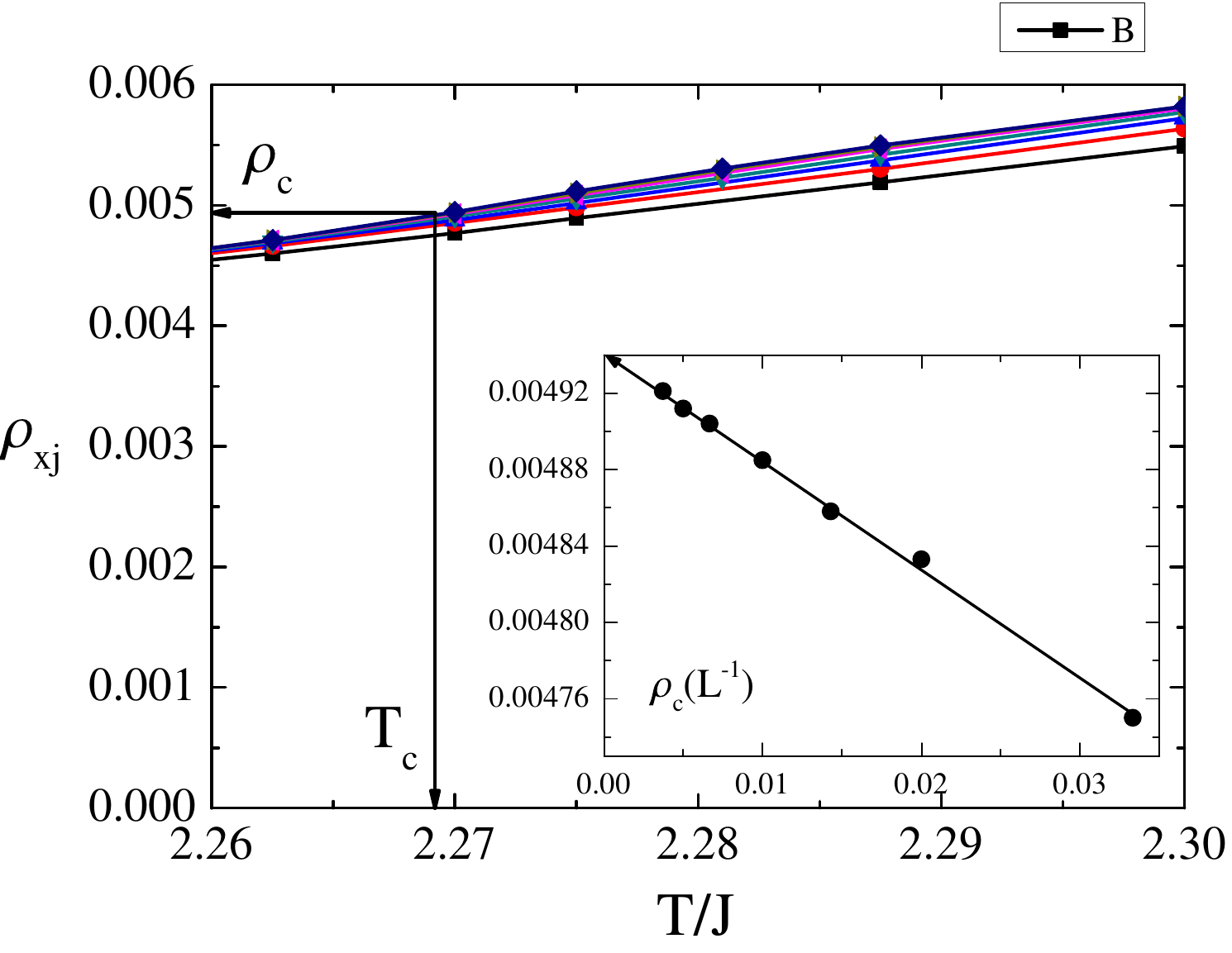}
b)
\includegraphics[scale=0.35]{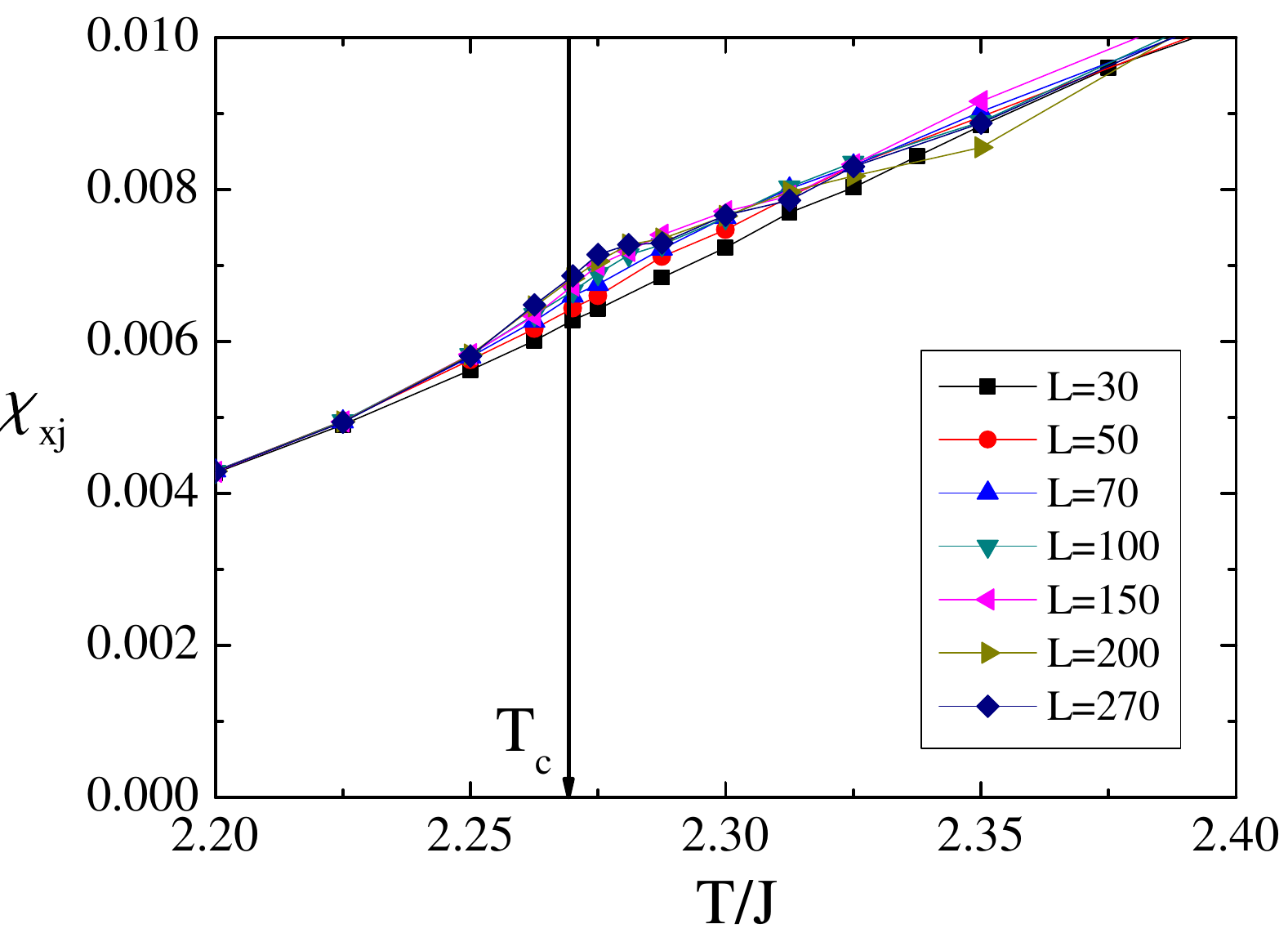}
\caption{\label{fig3} Wall X-junction density (a) and topological susceptibility (b) in the Ising model on a square lattice.}
\end{figure}%

Note that the topological susceptibility for X-junctions has no singularity at the critical temperature (fig. \ref{fig3}). So these defects are not relevant to the Ising transition. However, we expect that Y- and X-junctions become relevant in the q-state Potts model with $q\geq3$.

\section{$O(2)$ model}

Another simple case is the $O(2)$ model describing XY ferromegnets with $N=2$ (planar) spins. As well as the Ising model, this model is very representative,
since it is the simplest model with topological defects of another type and with another type of a phase transition, namely point-like defects and an infinite order transition.

The Hamiltonian of the XY model is
\begin{equation}
  H=-J\sum_{ij} \bfs_{i}\bfs_{j},\quad \bfs_{i}=(\cos\varphi_i,\sin\varphi_i).
\end{equation}
The order parameter space is $G/H=SO(2)$. Being continuous, such a symmetry cannot be spontaneously broken in two dimensions at nonzero temperature by local interactions, according to the Mermin-Wagner theorem. Since $\pi_1(SO(2))=\mathbb{Z}$, point-like defects (vortices) are presented in the system, effective long-range (logarithmical) interaction of which leads to the appearance of a quasi-long-range (algebraical) order below some temperature $T_\mathrm{v}$ by the BKT mechanism \cite{Berezinsky71,Berezinsky72,Kosterlitz73}.

As we have noted above, properties of a BKT transition can be obtained by the RG approach \cite{Kosterlitz73,Kosterlitz74}. One of the most important quantities characterizing a BKT transition is the spin stiffness (or the helicity modulus). The helicity modulus is defined by the increase in the free-energy density $F$ due to a small twist $\Delta_\mu$ across a system in one direction $\mu=1,2$:
\begin{equation}
\Upsilon_\mu=\left.\frac{\partial^2F}{\partial\Delta_\mu^2}\right|_{\Delta_\mu=0}
\end{equation}
In more detail, the helicity modulus is
\begin{equation}
    \Upsilon_\mu=\frac{J}{L^2}\left\langle\sum_{i<j}(\bfx_{ij}\cdot\bfe_\mu)^2\bfs_i\bfs_j\right\rangle -\frac{J^2}{L^2T}\left\langle\left(\sum_{i<j}(\bfx_{ij}\cdot\bfe_\mu)\bfs_i\times\bfs_j\right)^2\right\rangle,
    \label{U=u-x2}
\end{equation}
where $\bfx_{ij}$ is a vector between two sites, $\bfe_\mu$ is the direction of the twist.

\begin{figure}[t]
\center
a)
\includegraphics[scale=0.34]{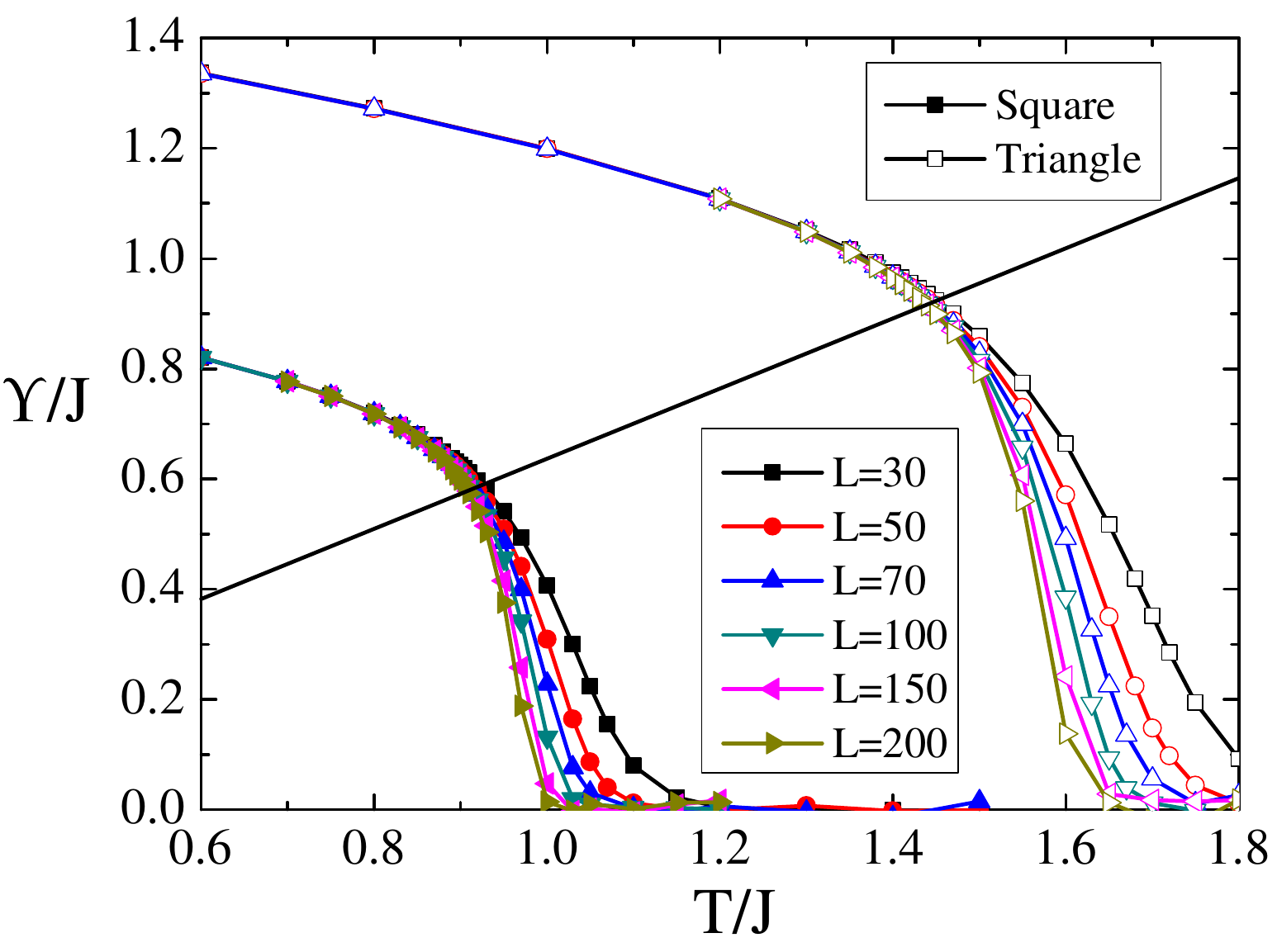}
b)
\includegraphics[scale=0.34]{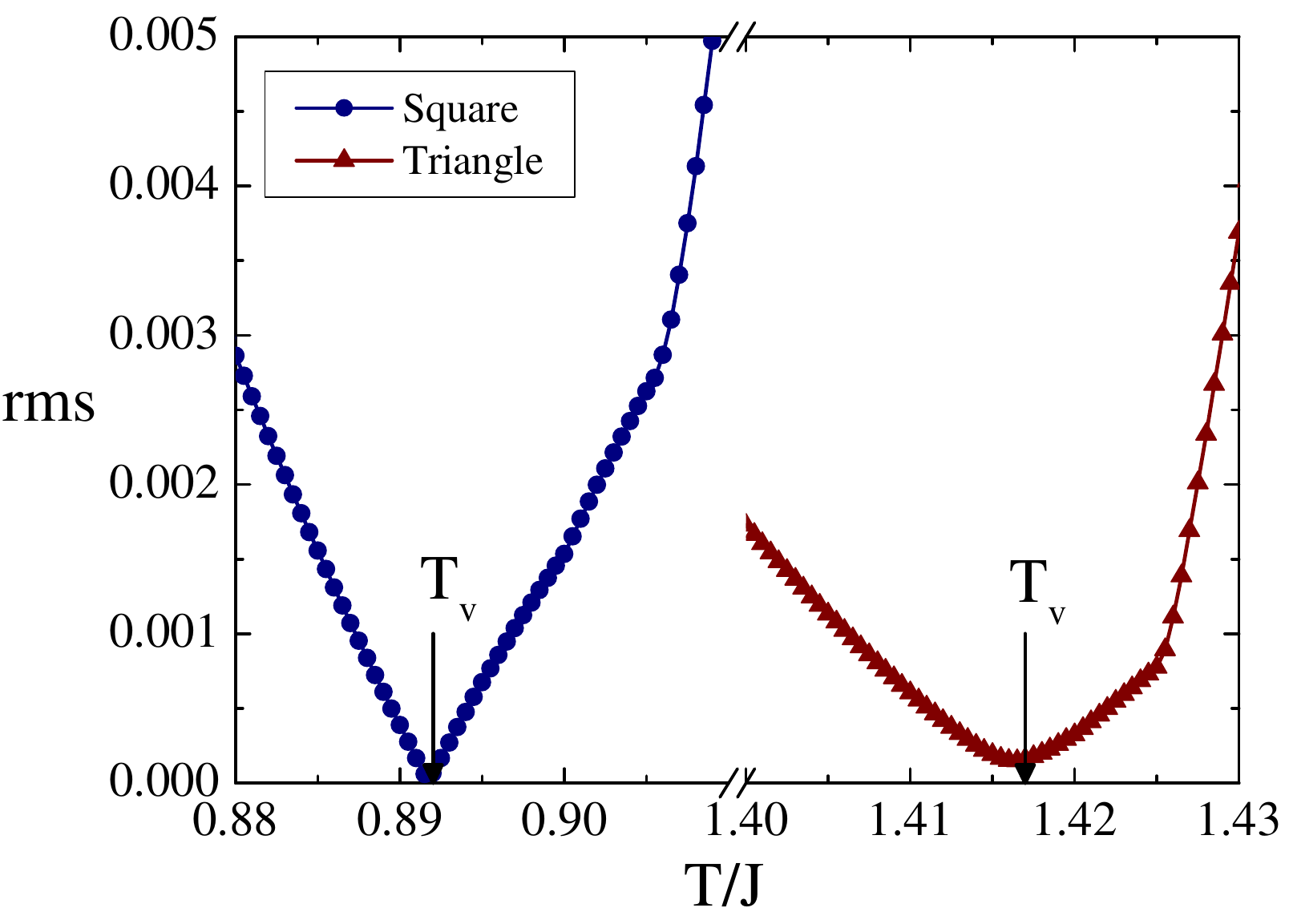}
\caption{\label{fig4} Helicity modulus (a) and the root-mean-square error of its fit near the BKT transition (b) in the $O(2)$ model on square and triangular lattices.}
\end{figure}

Important universal properties of a BKT transition predicted by the RG approach \cite{Nelson77} is the jump of the helicity modulus from $\frac{2T_\mathrm{v}}{\pi}$ to 0 at the transition point $T=T_\mathrm{v}$ (see fig. \ref{fig4}a). This property has become standard method to estimate the transition temperature. More precisely, in our simulations we use the Weber-Minnhagen finite-size scaling method \cite{Weber88} based on RG-corrections to the modulus $\Upsilon$
\begin{equation}
    \Upsilon(T,L)=\frac{2T}{\pi}\left(1+\frac1{2\ln L+c}\right),
    \label{weber}
\end{equation}
where fitting constant $c$ is selected so that the root-meansquare error $\mathrm{rms}(T)$ of the least-squares fit for formula (\ref{weber}) is minimal for each values of temperature (fig. \ref{fig4}b). The global minimum of $\mathrm{rms}(T)$ corresponds to the BKT transition temperature.

To study numerically the XY model, we use again the Wollf cluster algorithm \cite{Wollf89-2} and consider square and triangular lattices with sizes $L=30,$ 50, 70, 100, 150, and 200. The transition temperatures are estimated as ($J=1$)
\begin{equation}
    T_\mathrm{v}(\square)=0.892(1),\quad T_\mathrm{v}(\triangle)=1.418(2).
    \label{T-BKT}
\end{equation}
\begin{figure}[t]
\center
a)
\includegraphics[scale=0.35]{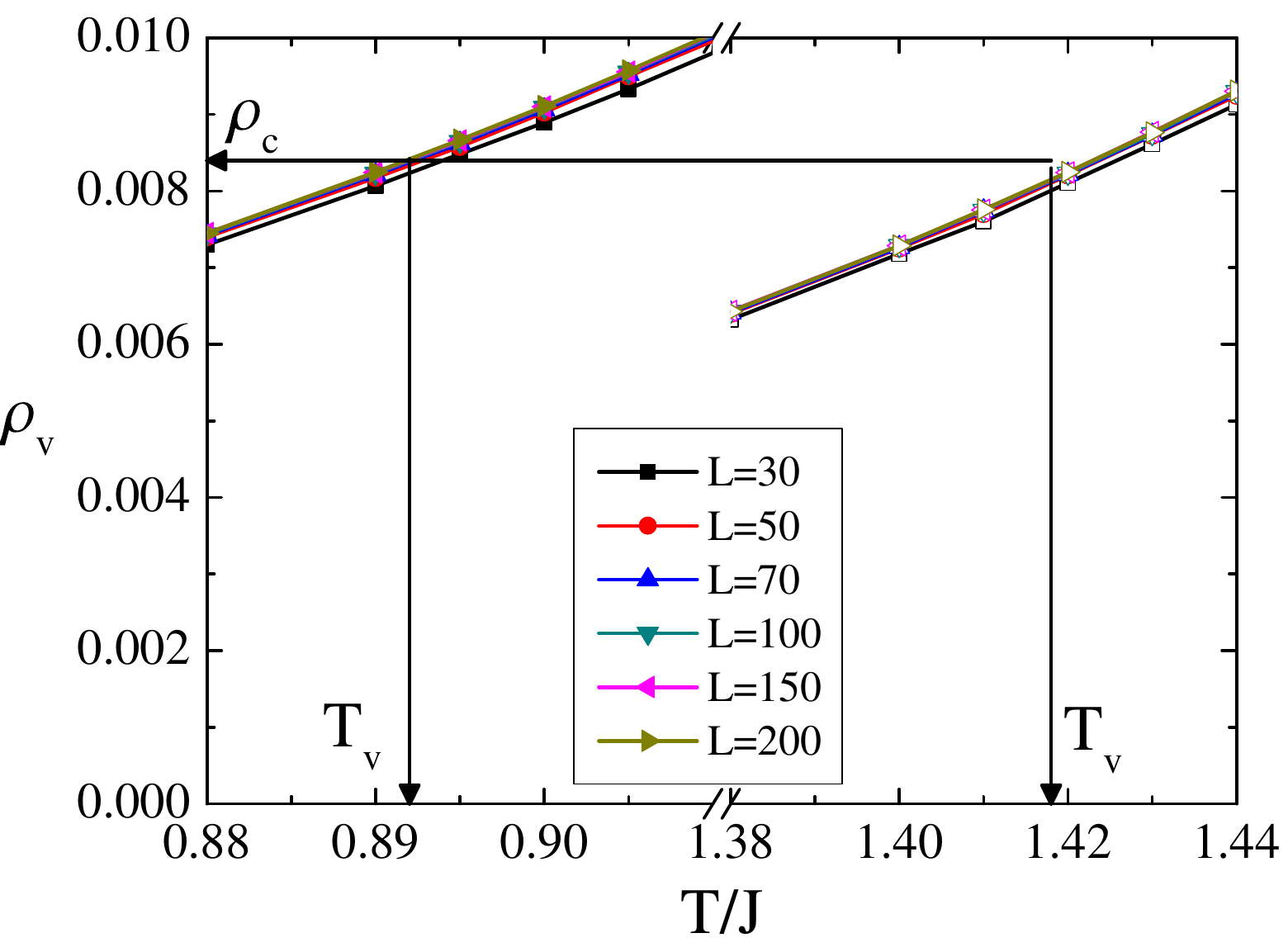}
b)
\includegraphics[scale=0.35]{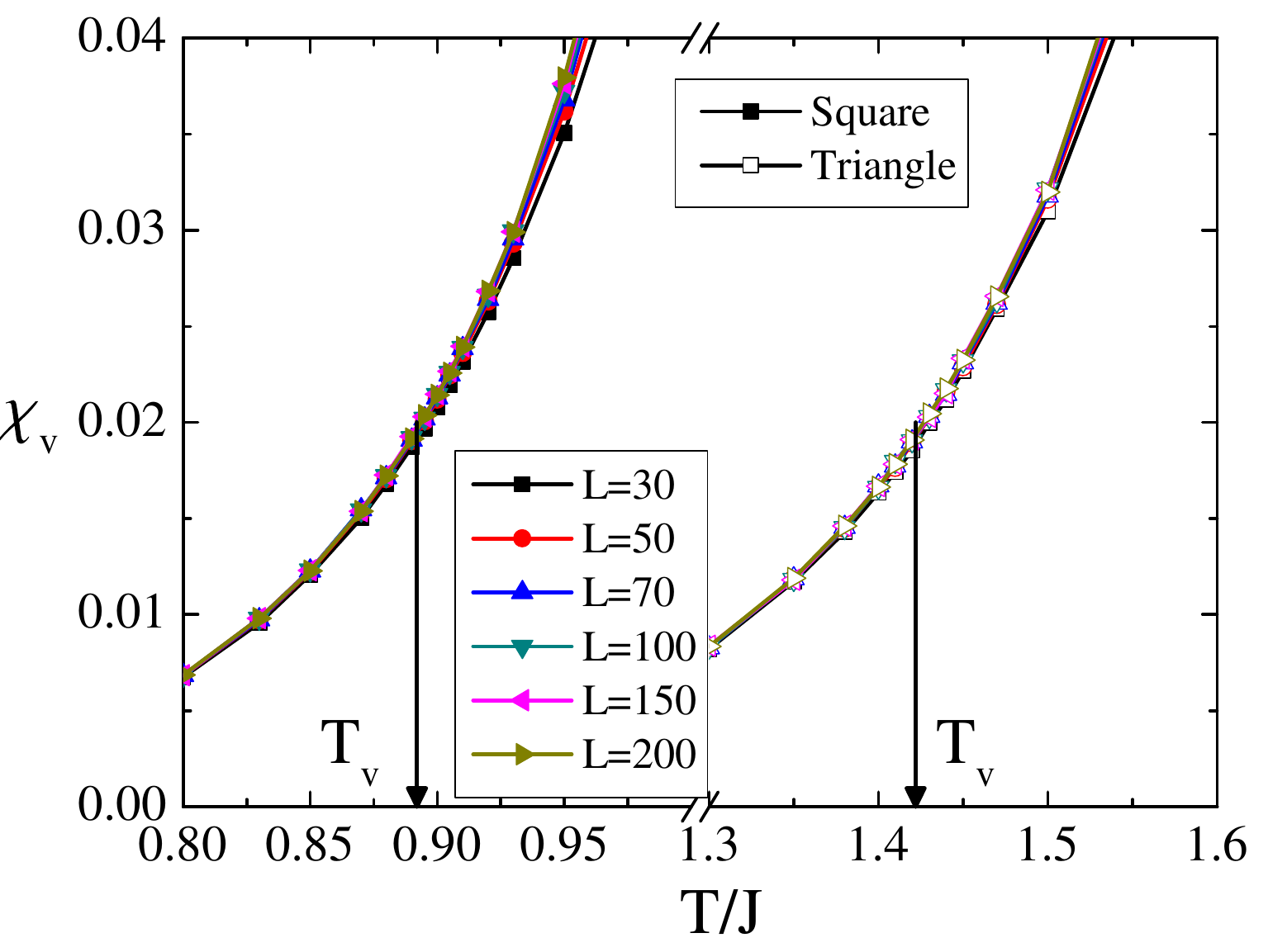}
\caption{\label{fig5} Vortex density (a) and topological susceptibility (b) in the $O(2)$ model on square and triangular lattices.}
\end{figure}%
The density of vortices is
\begin{equation}
    \tilde\rho_\mathrm{v}=\frac{1}{2\pi L^2}\sum_{x}\sum_{\square_x}\varphi_{ij},
    \quad \rho_\mathrm{v}=\langle\tilde\rho_\mathrm{v}\rangle,
\end{equation}
where $x$ runs over primitive cells of a lattice, and $\square_x$ means coming over a cell and summing differences of spin phases $\varphi_{ij}=\varphi_i-\varphi_j\in(-\pi,\pi]$ (remind that $\mathbf{S}=(\cos(\varphi_i),\sin(\varphi_i))$). The critical values of the density are (fig. \ref{fig5}a)
\begin{equation}
     \rho_\mathrm{v}(\triangle)=0.0083(3), \quad \rho_\mathrm{v}(\square)=0.0085(3),
\end{equation}
in good agreement with each other, but they are not consistent with the extrapolation of Halperins formula (\ref{r=a-eta}) with $\eta=1/4$ and $A=1/(2\pi)$: $\rho_\mathrm{v}\approx0.0177$, and with results of the procedure \cite{Antunes01} with $\rho_\mathrm{v}(T\to\infty)=0.333$: $\rho_\mathrm{v}\approx0.0740$.

Note that the topological susceptibility defined analogously to (\ref{chi}) (see fig. \ref{fig5}b) has no singularity at the transition temperature. Such a regular behavior of thermodynamical quantities is specific for infinite order transitions like the BKT one.

\section{$V_{3,2}$ Stiefel model}

As a non-trivial case, we consider the model where the order parameter space is $G/H=SO(3)$. Such a symmetry breaking scenario is related to frustrated magnetic system, namely it appears in magnets with a non-collinear but planar spin ordering and isotropic spins ($N=3$), e.g. frustrated helimagnets and antiferromagnet on a triangular lattice. A planar spin ordering is described by two orthogonal vectors. In general, a set of orientations of $P$ orthogonal $N$-vectors is the Stiefel manifold
\begin{equation}
    V_{N,P}=\frac{O(N)}{O(N-P)}.
    \label{manifold}
\end{equation}
In particular, $V_{3,2}=SO(3)$. The order parameter is a $3\times 2$ matrix $\Phi$ composed of two orthogonal unit $3$-vectors
\begin{equation}
    \Phi\left(V_{3,2}\right)=(\bfs,\bfk).
\end{equation}
The simplest model with $G/H=SO(3)$ is a natural generalization of the Heisenberg model \cite{Zumbach93}
\begin{equation}
    H=-J\sum_{ij}\tr\,\Phi_i^T\Phi_j.
    \label{Stiefel}
\end{equation}
As far as $\pi_1(SO(3))=\mathbb{Z}_2$, the model contains topological defect of a new type, so-called $\mathbb{Z}_2$ vortices, dissimilar to usual $\mathbb{Z}$ vortices of the XY model. The first difference is that $\mathbb{Z}_2$ vortices and $\mathbb{Z}_2$ antivortices are the same, so any two vortices annihilate each other. Association of pairs vortex-antivortex increases ordering of a system, but it doesn't lead to appearance of long-range or quasi-long-range orders. Another difference is technical one: $SO(3)$ group is non-Abelian, so perturbative excitations (spin waves) can not be integrated out unlike to $SO(2)$ Abelian case, and we can not obtain clear picture of a vortices interaction.

Nevertheless, $\mathbb{Z}_2$ vortices have an interesting "almost critical"\ behavior. Point is that the increasing of temperature and concentration of vortices (not associated in pairs) leads to a rather sharp change in the temperature behavior. At low temperatures, when the vortex density is small, the system behaves in according with the prediction of the $O(4)$ sigma model \cite{Azaria90,Azaria93}, and after the crossover the behavior changes to some high-temperature one. (For a brief historical review of research on $\mathbb{Z}_2$ vortices see \cite{Sorokin17}.) If one considers the vortex density as a critical parameter, the crossover is reminiscent distantly of a crossover in a supercritical fluid in a liquid-gas phase diagram. To reach a possible critical point, it is necessary to have the chemical potential of vortices as a free parameter, that is absent in the original model. But the presence of the pronounced crossover means that the original model is closed to the critical point, so one can talk about an "almost critical"\ behavior. Moreover, as we show recently for the Ising-$V_{3,2}$ (or $V_{3,3}$) model \cite{Sorokin17}, the coinciding crossover and Ising transition become the first order transition, and therefore $\mathbb{Z}_2$ vortices are significant participants of the critical phenomena. Anyway, we assume that the $\mathbb{Z}_2$ vortex density has an almost universal value at a crossover temperature.

\begin{figure}[t]
\center
a)
\includegraphics[scale=0.34]{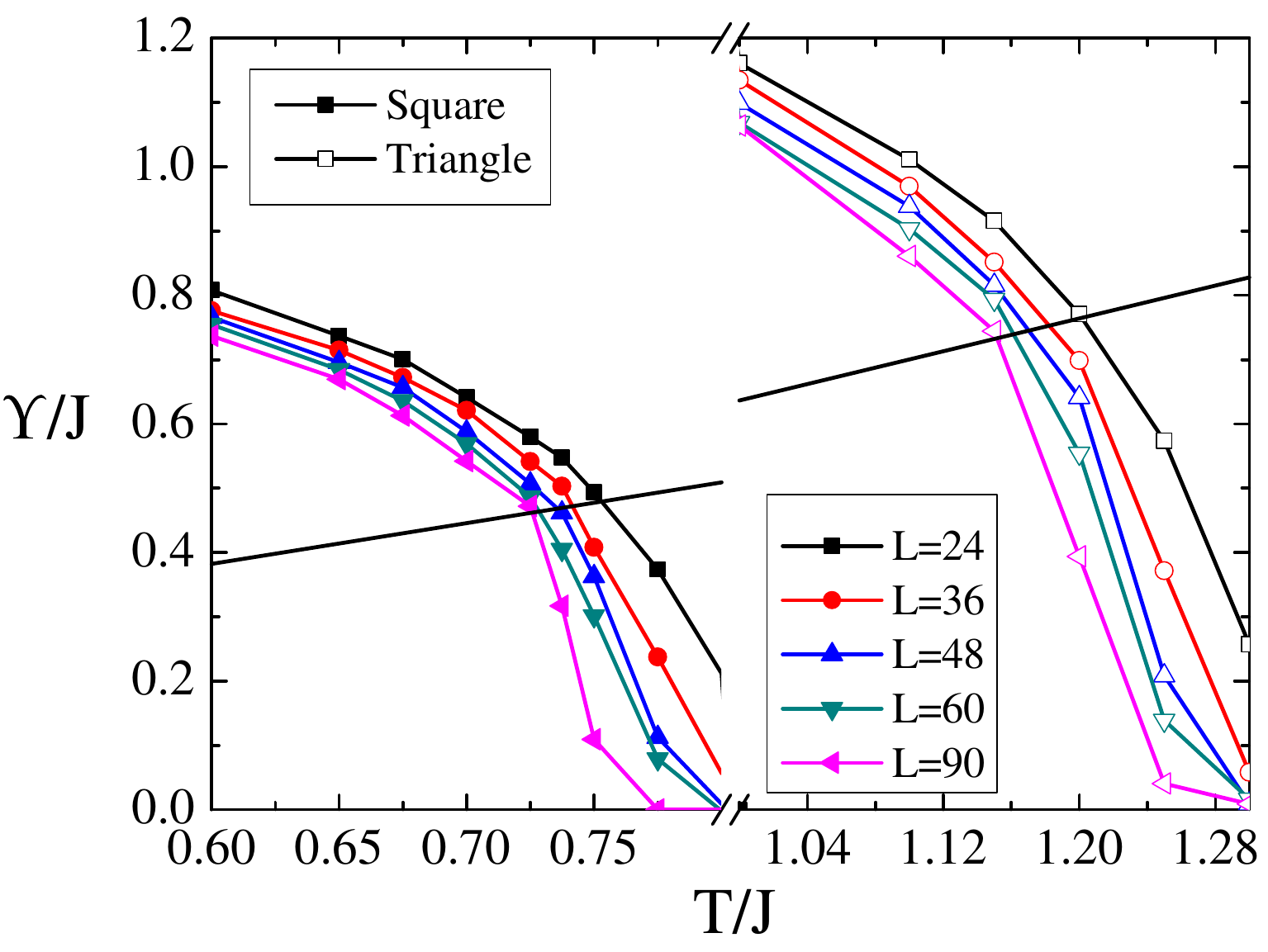}
b)
\includegraphics[scale=0.34]{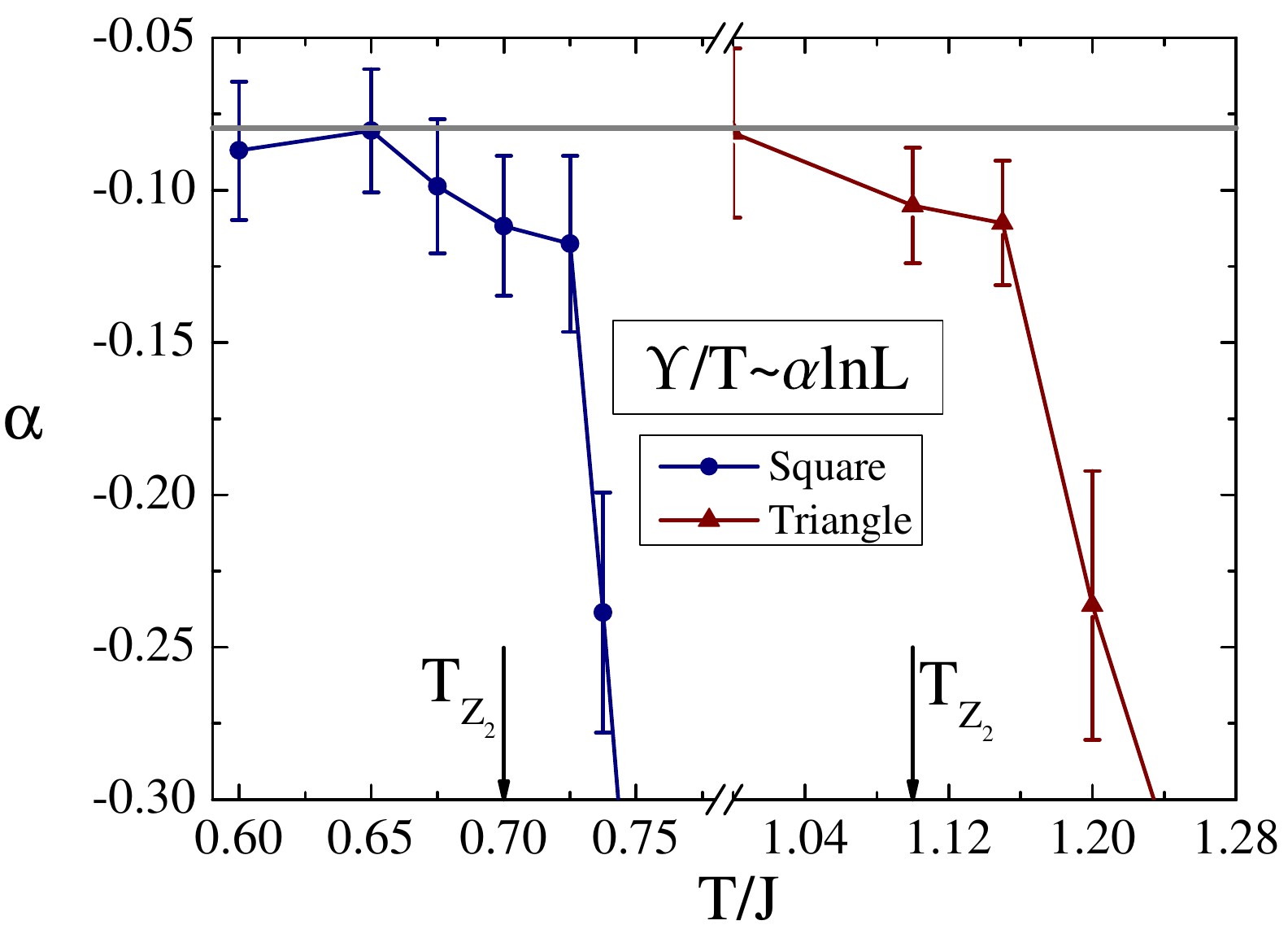}
\caption{\label{fig6} Helicity modulus (a) and the slope of the linear fit (\ref{azaria}) (b) in the $V_{3,2}$ Stiefel model on square and triangular lattices.}
\end{figure}%
To define the vortex density, one should extend the order parameter $\Phi$ to a $3\times3$ orthogonal matrix by adding the third vector $\bft=\bfs\times\bfk$, so that $\det\Phi=1$, then
\begin{equation}
    \tilde\rho_{\mathbb{Z}_2}=\frac{1}{2L^2}\sum_x \left(1-\frac12\tr\prod_{\square_x} f(\Phi_i^{-1}\Phi_j)\right),\quad
    \rho_{\mathbb{Z}_2}=\langle\tilde\rho_{\mathbb{Z}_2}\rangle,
\end{equation}
where $f:SO(3)\to SU(2)$ is homomorphism describing double covering of $SO(3)$ by $SU(2)$ (in practice, an unitary matrix can be constructed using the parametrization of corresponding orthogonal matrix by Euler angles). The crossover temperature is determined as a point where the temperature dependence of the correlation length changes its behavior from the $O(4)$ sigma model regime to the high-temperature one. The RG calculation predicts \cite{Azaria92} that in the $O(4)$ sigma model the helicity modulus behaves as
\begin{equation}
    \frac{\Upsilon(L)}{T}\sim\frac1{4\pi}\ln\left(\frac{\xi}{L}\right).
    \label{azaria}
\end{equation}
The helicity modulus is calculated using the obvious extension of formula (\ref{U=u-x2}) to the case of two $N=3$ spin $\mathbf{V}^\alpha=\bfs,\bfk$; $\alpha=1,2$; $\mathbf{V}=(V^1,V^2,V^3)$:
$$
    \Upsilon_\mu^a=\frac{J}{L^2}\left\langle\sum_{i<j,\alpha}(\bfx_{ij}\cdot\bfe_\mu)^2(V_i^{b,\alpha}V_j^{b,\alpha}+V_i^{c,\alpha}V_j^{c,\alpha})\right\rangle-
$$
\begin{equation}
    -\frac{J^2}{L^2T}\left\langle\left(\sum_{i<j,\alpha}(\bfx_{ij}\cdot\bfe_\mu)(V_i^{b,\alpha}V_j^{c,\alpha}-V_i^{c,\alpha}V_j^{b,\alpha})\right)^2\right\rangle,
    \label{U2=u-x2}
\end{equation}
\begin{equation}
    \Upsilon_\mu=\frac13\sum_{a=1}^3\Upsilon_{\mu,a},
\end{equation}
To study the $V_{3,2}$ models, we use Monte Carlo simulations based on the over-relaxed algorithm \cite{Brown87,Creutz87}. Lattice sizes are $L=24,\,36,\,48,\,60,$ and 90. Thermalization is performed within $3\cdot10^5$ Monte Carlo steps per spin, and calculation of averages within $2.4\cdot10^6$ steps. Previous results for the case on a square lattice have been published in \cite{Sorokin17}.

\begin{figure}[t]
\center
a)
\includegraphics[scale=0.35]{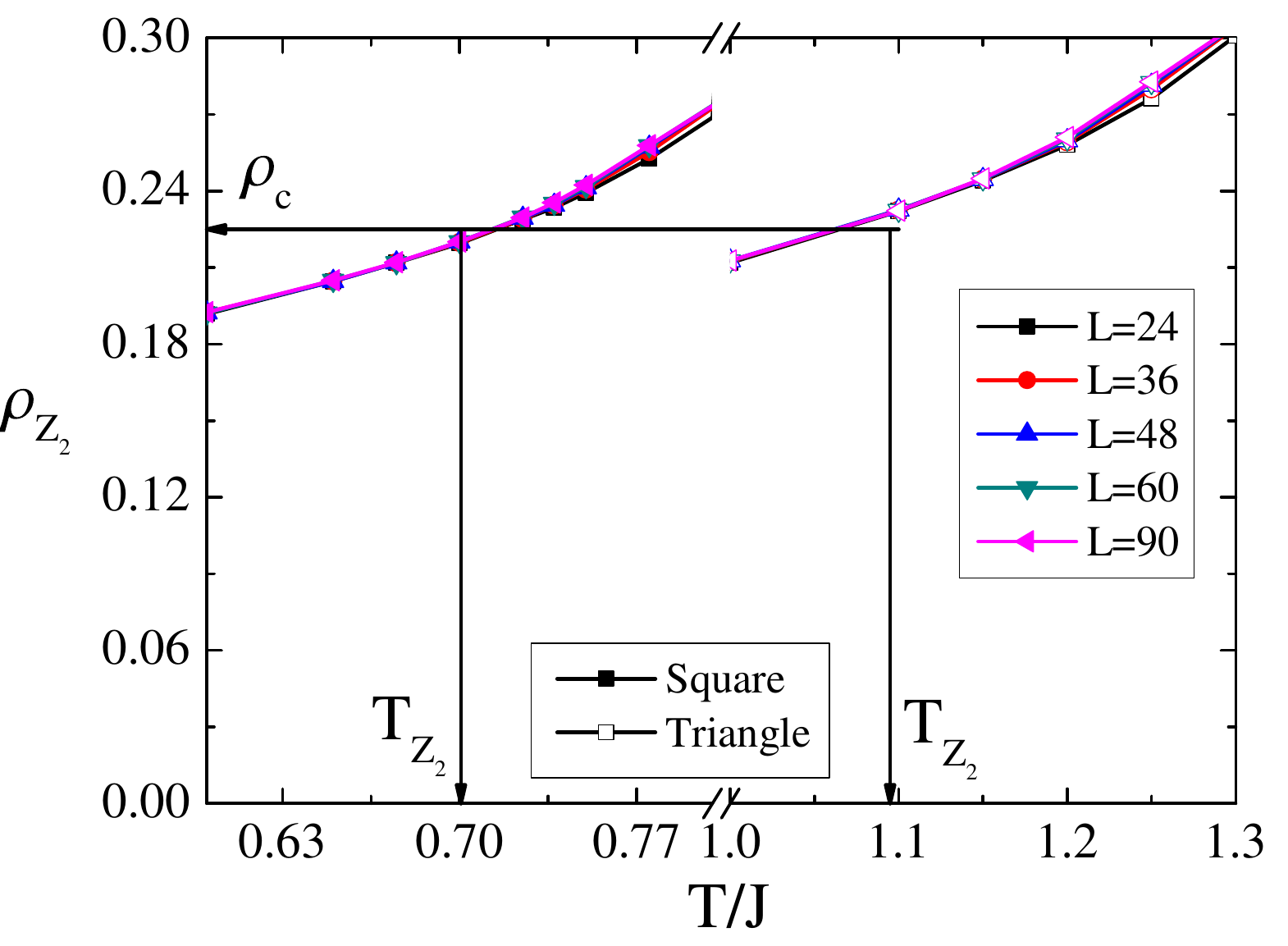}
b)
\includegraphics[scale=0.35]{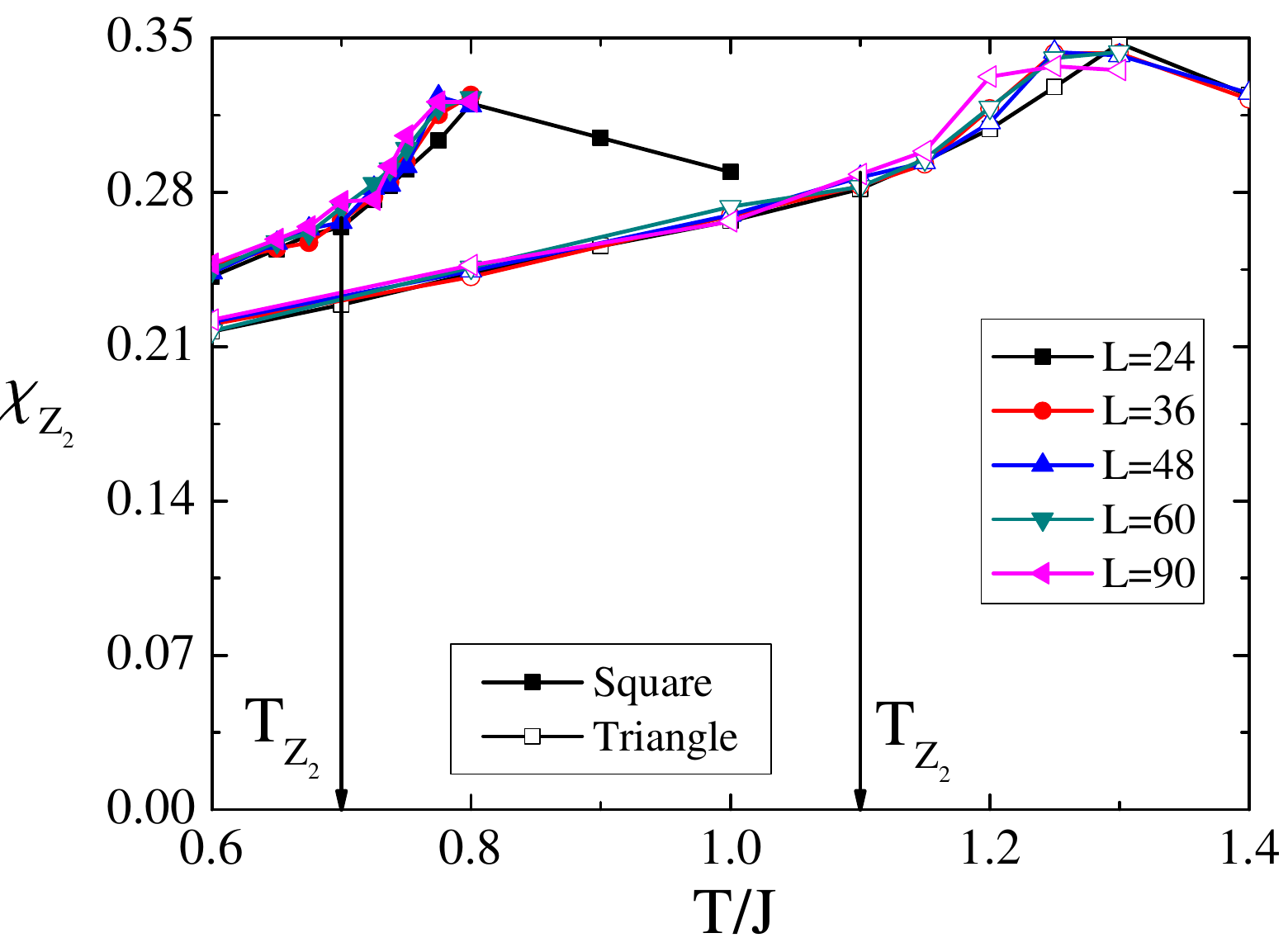}
\caption{\label{fig7} Vortex density (a) and topological susceptibility (b) in the $V_{3,2}$ model on square and triangular lattices.}
\end{figure}%
The crossover temperatures are estimated as (fig. \ref{fig6})
\begin{equation}
    T_{\mathbb{Z}_2}(\square)=0.70(2),\quad T_{\mathbb{Z}_2}(\triangle)=1.10(3).
\end{equation}
The density of $\mathbb{Z}_2$ vortices is (fig. \ref{fig7}a)
\begin{equation}
     \rho_{\mathbb{Z}_2}(\triangle)=0.221(10), \quad \rho_{\mathbb{Z}_2}(\square)=0.229(15).
\end{equation}
Thus, the values of the vortex density at the crossover temperature coincide within the limits of the error. Note that the main error arises from an estimate of the crossover temperature.

Similar to a BKT transition, the topological susceptibility has no singularity at the crossover temperature (fig. \ref{fig7}b).

\section{Ashkin-Teller model}

The universality of the critical behavior at a continuous transition point implies primarily independence of parameters describing this behavior on details at the microscopic level, as long as a symmetry remains the same. I.e. on a large scale, a system must cease to perceive a structure of a lattice as well as all infrared-insignificant interactions and phenomena, which do not change a symmetry of an effective model. In particular, an additional type of topological defects may be presented in a system, and if these defects are not relevant to a phase transition, they must not affect the critical behavior at first sight. In fact, the situation is some more delicate. One knows examples where two types of topological defects are represented in a system, and their interaction affects the temperature, sequence and type of phase transitions. Moreover, an interaction between two types of defects may lead to appearance of an extra type of defects. Such an example is discussed in the next section. So existence of additional type defects is even more reliable test of our hypothesis (that the critical value of the topological defect density is universal) than comparison results for different lattices.

The simplest model with a defect-defect interaction is the Ashkin-Teller (AT) model \cite{Ashkin43}. This model is equivalent to two interacting Ising models \cite{Fan72} with the Hamiltonian
\begin{equation}
    H=-\sum_{ij}\left(J(s_i s_j+\sigma_i\sigma_j)+J_4s_i s_j\sigma_i\sigma_j\right),\quad s_i,\sigma_i=\pm1.
  \label{At-model}
\end{equation}
Let us remind some facts about the phase diagram of the two-dimensional AT model (see, e.g., \cite{Kadanoff80}). At $J_4/J<-1$, spins $s$ and $\sigma$ are individually disordered $\langle s\rangle=\langle\sigma\rangle=0$, but the system has the antiferromagnetic order in the parameter $\langle s\sigma\rangle$, which is destroyed above the temperature of the single Isingian phase transition. At $-1<J_4/J\leq1$, the ordered phase is described by three order parameters $\langle s\rangle$, $\langle \sigma\rangle$, $\langle s\sigma\rangle\neq0$. It is co-called Baxter phase. In this case, one observes a single second order phase transition to the disordered phase with continuously varying critical exponents \cite{Kadanoff79,Kadanoff792}. In particular, the case $J_4/J=0$ corresponds to two decoupled Ising models, and the case $J_4/J=1$ is the 4-state Potts model. The possibility of continuous varying of critical indices is realized at two-dimensional critical points described by the conformal field theory with the central charge $c\geq1$ \cite{Zamolodchikov86,Zamolodchikov87}. The critical line of the AT model at $-1<J_4/J\leq1$ corresponds to the case $c=1$. Especially since the estimation of the fractal dimension of walls (as one of critical exponents) is varying too in dependence on $J_4/J$ \cite{Caselle11,Ikhlef12}, we don't expect that the domain wall density has an universal critical value along this critical line. So, for example, $\rho_\mathrm{dw}=0.2461(2)$ for the 4-state Potts model instead of $\rho_\mathrm{dw}=2\times0.1464$ for two decoupled Ising models.

At $J_4/J>1$, two alternate Isingian phase transitions occur with increasing temperature. The first transition is from the Baxter phase $\langle s\rangle\neq0$, $\langle s\sigma\rangle\neq0$ to the partially ordered phase with $\langle s\sigma\rangle\neq0$, and the second one is to the disordered phase. We are interested in these transitions.

We investigate one case of the AT model with $J=1$ and $J_4=2$ using Monte Carlo simulations based on the Metropolis algorithm. We consider lattices with sizes $L=20,\,30,\,50,\,70,\,100$ and 150.

\begin{figure}[t]
\center
a)
\includegraphics[scale=0.34]{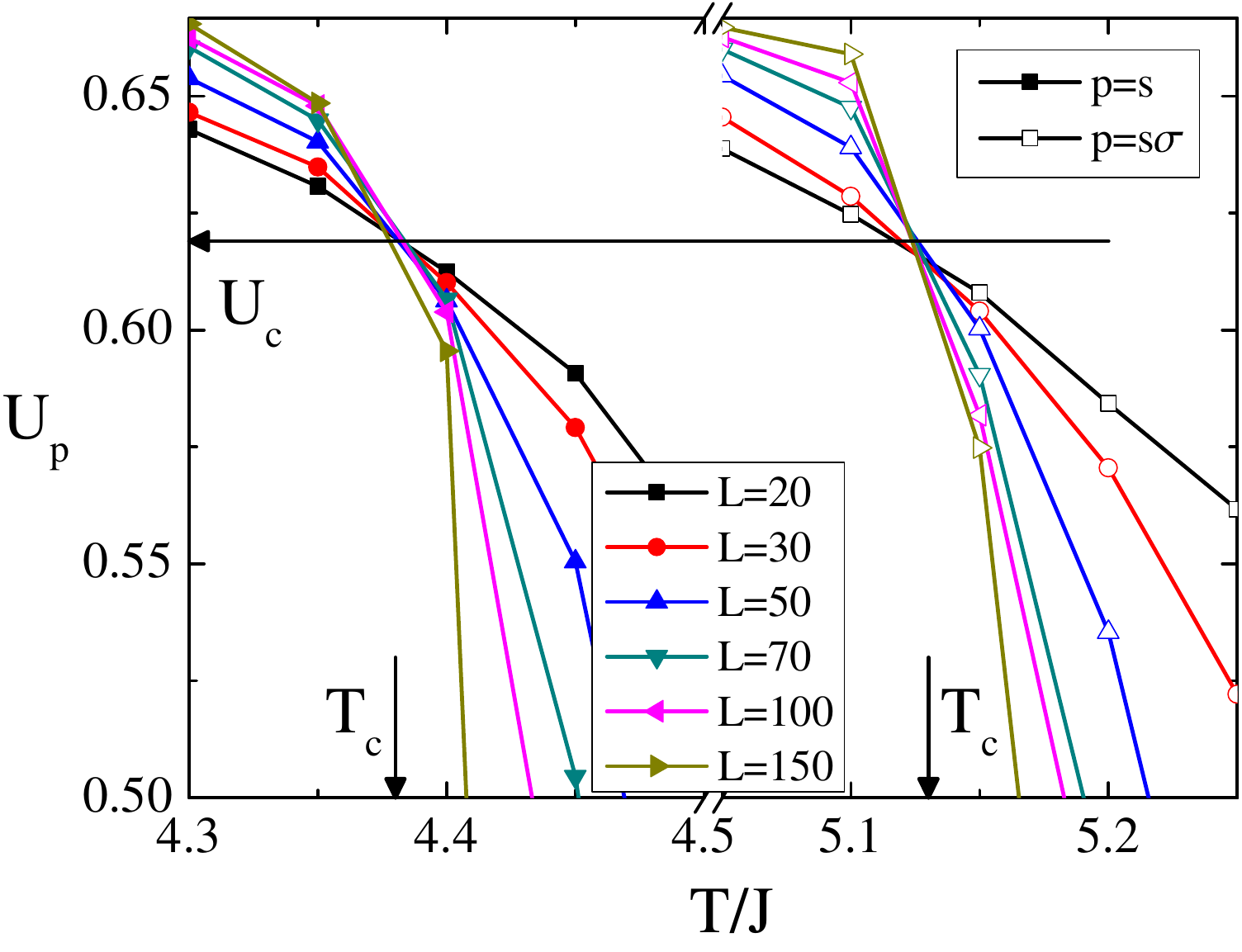}
b)
\includegraphics[scale=0.34]{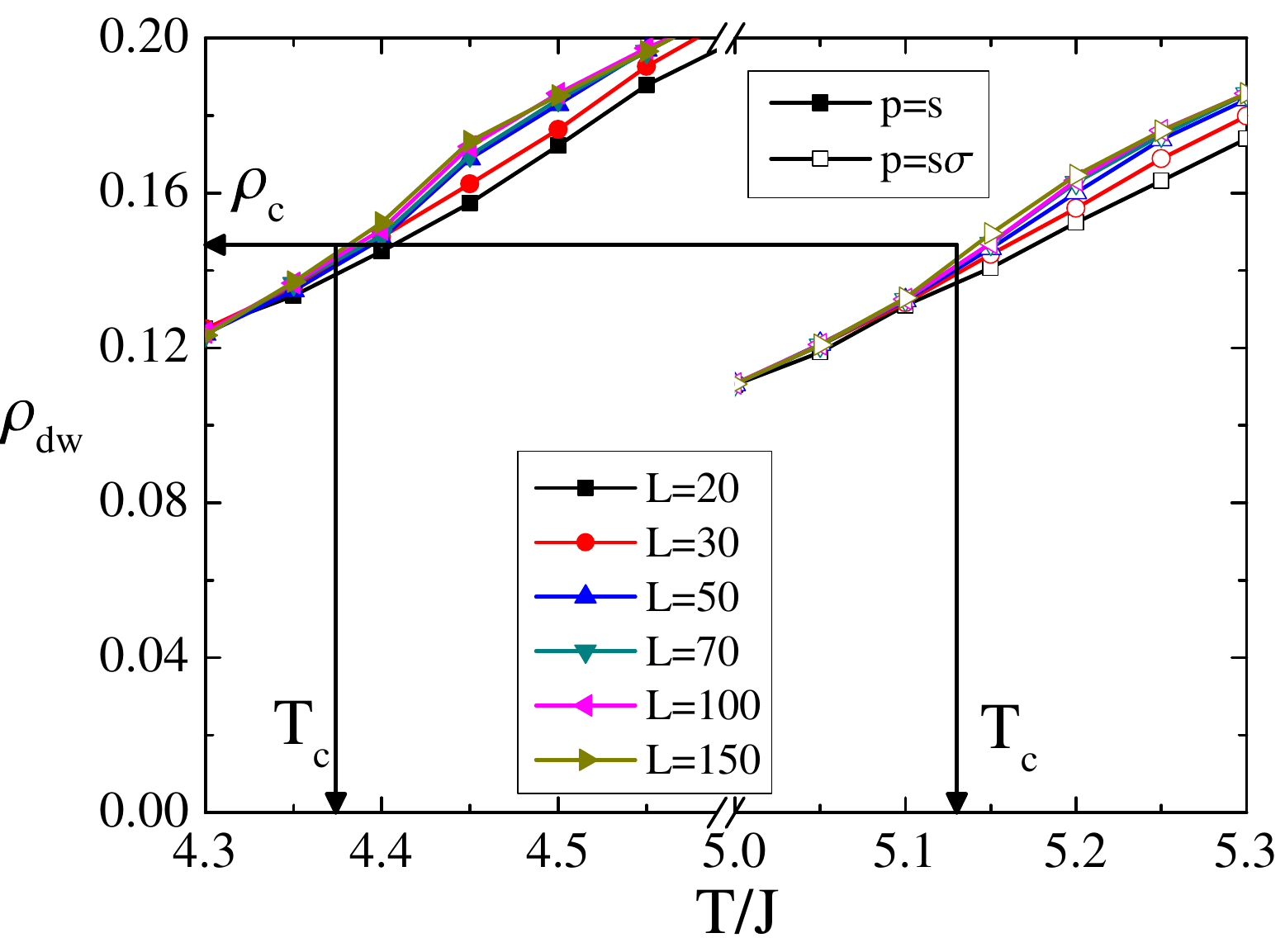}
\caption{\label{fig8} Binder cumulants (a) and wall density (b) in the AT model on a square lattice.}
\end{figure}%
In contrast to the case $-1<J_4/J\leq1$ where the transition temperature is known via a duality statement, in our case we estimate the critical temperatures numerically by the Binder cumulant crossing method \cite{Binder81}. The cumulants are
\begin{equation}
    U_p=1-\frac{\langle p^4 \rangle}{3\langle p^2 \rangle^2},
\end{equation}
where $p$ is an order parameter corresponding to a transition. At the first transition, the order parameter is
\begin{equation}
    p=s=\frac{1}{L^2}\left|\sum_i s_i\right|,
\end{equation}
at the second one,
\begin{equation}
    p=s\sigma=\frac{1}{L^2}\left|\sum_i s_i\sigma_i\right|.
\end{equation}
For both order parameters, we introduce the domain wall density (\ref{ro-dw-ising}), the topological susceptibility (\ref{chi}) and the density of wall X-junctions (\ref{X-junction}).

The critical temperatures for a square lattice are estimated as (fig. \ref{fig8}a)
\begin{equation}
    T_\mathrm{dw}^s(\square)=4.374(2),\quad
    T_\mathrm{dw}^{s\sigma}(\square)=5.129(3),
\end{equation}
and for a triangular lattice
\begin{equation}
    T_\mathrm{dw}^s(\triangle)=6.944(4),\quad
    T_\mathrm{dw}^{s\sigma}(\triangle)=8.127(5).
\end{equation}
\begin{figure}[t]
\center
a)
\includegraphics[scale=0.35]{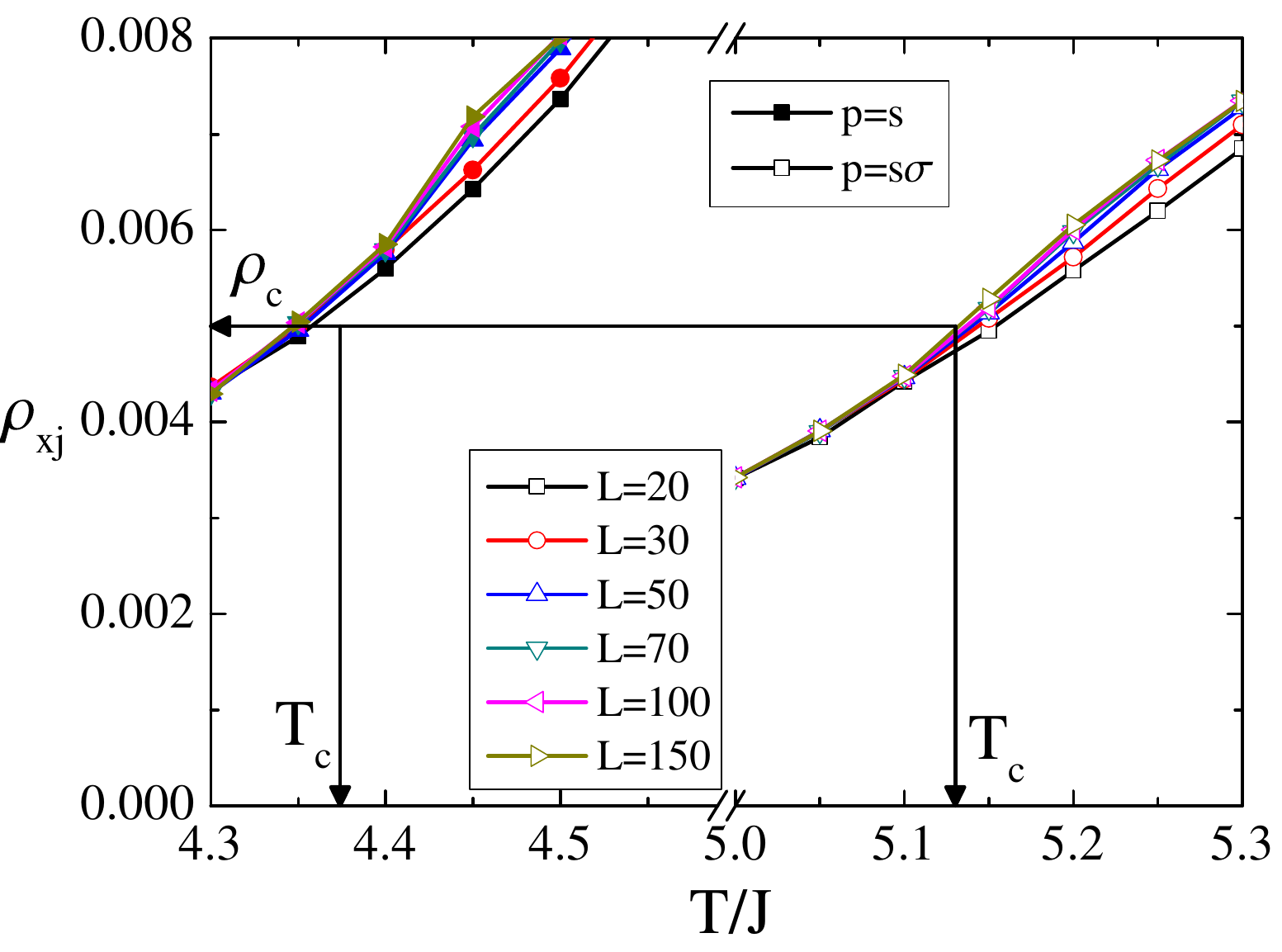}
b)
\includegraphics[scale=0.35]{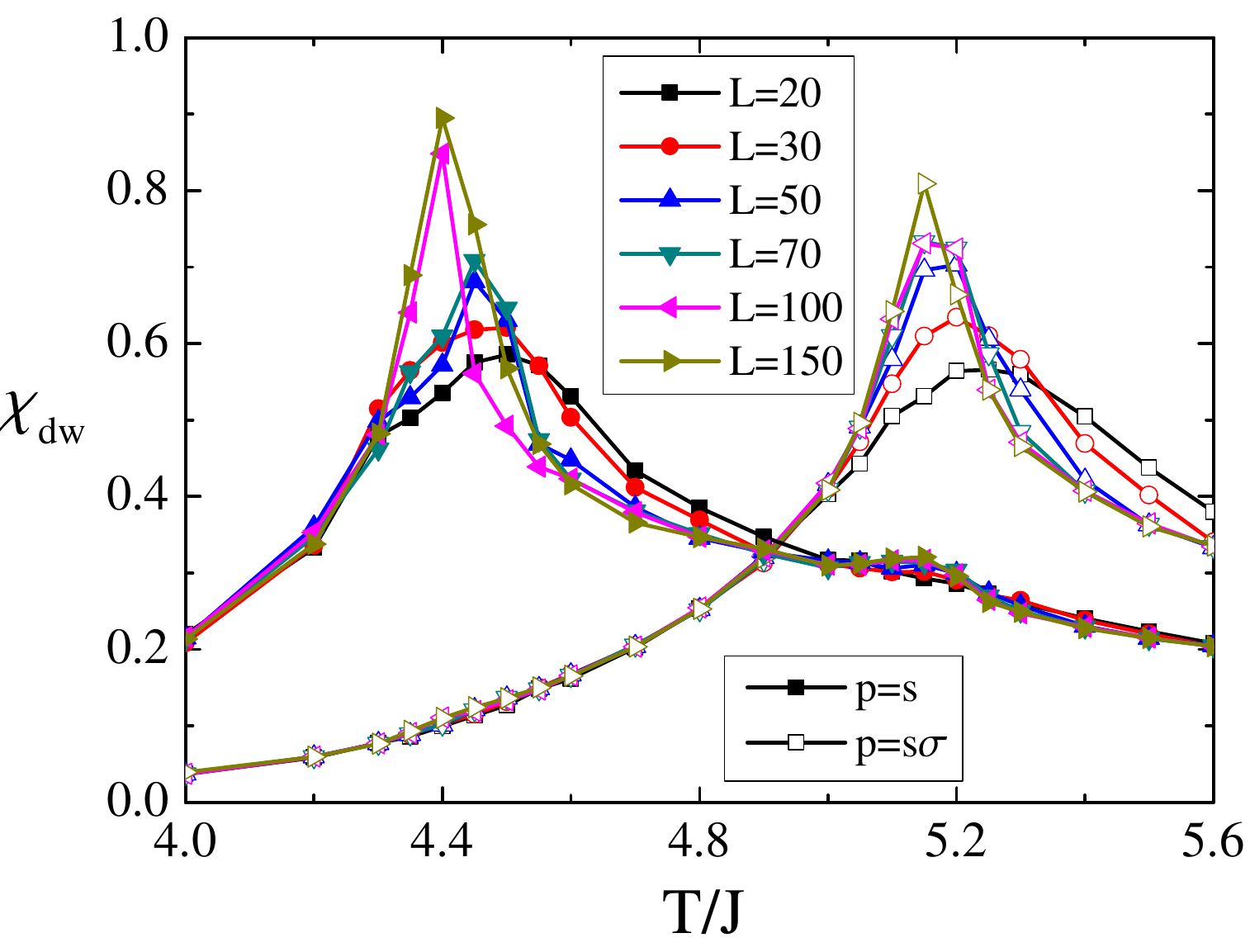}
\caption{\label{fig9} Walls junction density (a) and topological susceptibility (b) in the AT model on square lattice.}
\end{figure}%
The critical values of the wall and X-junction densities are (figs. \ref{fig8}b and \ref{fig9}a)
\begin{equation}
    \rho_\mathrm{dw}^s(\triangle)=0.167(1),\quad
    \rho_\mathrm{dw}^{s\sigma}(\triangle)=0.166(1),
\end{equation}
\begin{equation}
    \rho_\mathrm{dw}^s(\square)=0.1470(7),\quad
    \rho_\mathrm{dw}^{s\sigma}(\square)=0.1461(8),
\end{equation}
\begin{equation}
    \rho_\mathrm{xj}^s(\square)=0.0055(7),\quad
    \rho_\mathrm{xj}^{s\sigma}(\square)=0.0050(8).
\end{equation}
These results are in a good agreement with the values of the Ising model.

Note that the topological susceptibility corresponding to one parameter perceives a transition in another parameter (fig. \ref{fig9}b). This is manifested as a small growth of the susceptibility near a transition point, but such a growth does not seem to be a singularity in the thermodynamical limit $L\to\infty$. So, domain walls of two types  interact, but this interaction does not affect the critical behavior.

\section{Ising-$O(2)$ and $V_{2,2}$ Stiefel models}

The next simplest example of a model with a defect-defect interaction is the Ising-XY model \cite{Granato87,Granato91,Kosterlitz91}
\begin{equation}
  H=-\sum_{ij} \left((J+J_b\sigma_i\sigma_j)\bfs_{i}\bfs_{j}+J_c\sigma_i\sigma_j\right),\quad \bfs_{i}=(\cos\varphi_i,\sin\varphi_i), \quad \sigma_i=\pm1.
\end{equation}
There are two types of topological defects in this system: domain walls and vortices. This model and $G/H=O(2)\equiv\mathbb{Z}_2\otimes SO(2)$ symmetry class have a sizable number of physical realizations (see \cite{Korshunov06} for a review), including frustrated XY magnets with non-collinear spin ordering.

At $J_b=0$, the model is equivalent to decoupled Ising and $O(2)$ models. As long as the critical point of the Ising model corresponds to the conformal field theory with the central charge value $c=\frac12$, and the $O(2)$ model corresponds to $c=1$, simultaneous phase transition in both (discrete and continuous) order parameters is expected to be described by the $c=\frac32$ (super)conformal field theory. In the same way as the value $c=1$ of the central charge affiliates two interacting Ising models (i.e. $n_f=2$ AT model) with the free bosonic field, $O(2)$ and 4-state Potts  models, the value $c=\frac32$ affiliates the Ising-XY model with the free superfield \cite{Friedan85}, supersymmetric BKT transition \cite{Goldschmidt86,Doria86}, $O(3)$ Gross-Neveu, $n_f=3$ Ashkin-Teller \cite{Shankar85,Goldschmidt862}, and coupled 3-state Potts and tricritical Ising \cite{Dixon88} models. (The relations between the Ising-XY and other $c=\frac32$ models have been discussed in \cite{Foda88}.) The anisotropic $n_f=3$ Ashkin-Teller model has the rich phase diagram with non-trivial continuous transitions at least in a mean-field calculation, including critical lines with continuously varying exponents and Ising-BKT transition points \cite{Goldschmidt862}. However, the most accurate Monte Carlo simulations \cite{Hasenbusch05,Hasenbusch052} indicate that some non-trivial critical behavior is not realized in the Ising-XY model. Depending on values of the parameters, either two separate transitions occur in the system, or one of the first order. (A possible multicritical behavior is unknown.)

In this paper, we are only interested in two cases with well-separated transitions temperatures either $T_\mathrm{v}>T_\mathrm{dw}$ or $T_\mathrm{v}<T_\mathrm{dw}$. In the first case, we choose $J_c=0$. If $J_b=J$, one has $T_\mathrm{v}<T_\mathrm{dw}$, but the transition temperatures $T_\mathrm{v}$ and $T_\mathrm{dw}$ are close to each other. Generally speaking, the case $J_b=J$ at any value of $J_c$ has the special property that the Ising disorder induces the XY disorder because a domain wall makes XY spins decoupled at a wall: $J+J_b\sigma_i\sigma_j=0$. Below the value $J_c<-1$, the quasi-long-range order in XY spins is absent at non-zero temperatures. In the range of values $-1<J_c<J_c^{B}<0$ with some value of the bifurcation point $J_c^B$, the Ising and BKT transitions occur at the same temperature as a first order transition. At $J_c>J_c^B$, the Ising transition temperature is above the BKT transition.

When $J_c=J_b=0$, the BKT transition temperature has the usual value (\ref{T-BKT}), while the Ising transition has zero temperature, which starts to increase with increasing of $J_b$, and $T_\mathrm{v}>T_\mathrm{dw}$. So at some value $J_b^M<1$, one finds the multicritical (tetracritical) point, where the sequence of the phase transitions temperatures becomes inverse $T_\mathrm{v}<T_\mathrm{dw}$. At $J_c=0$, we find numerically $J_b^M/J\approx 0.70(2)$.

In the first case $T_\mathrm{v}>T_\mathrm{dw}$, we choose $J_b=0.25$ and $J=1$, in the second one $T_\mathrm{v}<T_\mathrm{dw}$, we choose $J_c=J_b=J=1$.

The study of the Ising-XY models has been performed analogously to the study of the $V_{3,2}$ Stiefel model. Note that the case $J_c=0$, $J_b=J$ of the Ising-XY model is equivalent to the $V_{2,2}$ Stiefel model (\ref{Stiefel}) with $\bfk=\sigma(-\sin\varphi,\cos\varphi)$. And the case $J_c=0$ and varying $J_b$ can be considered as the interpolating model between the $V_{2,2}$ and $V_{2,1}\equiv O(2)$ models.

\begin{figure}[t]
\center
a)
\includegraphics[scale=0.34]{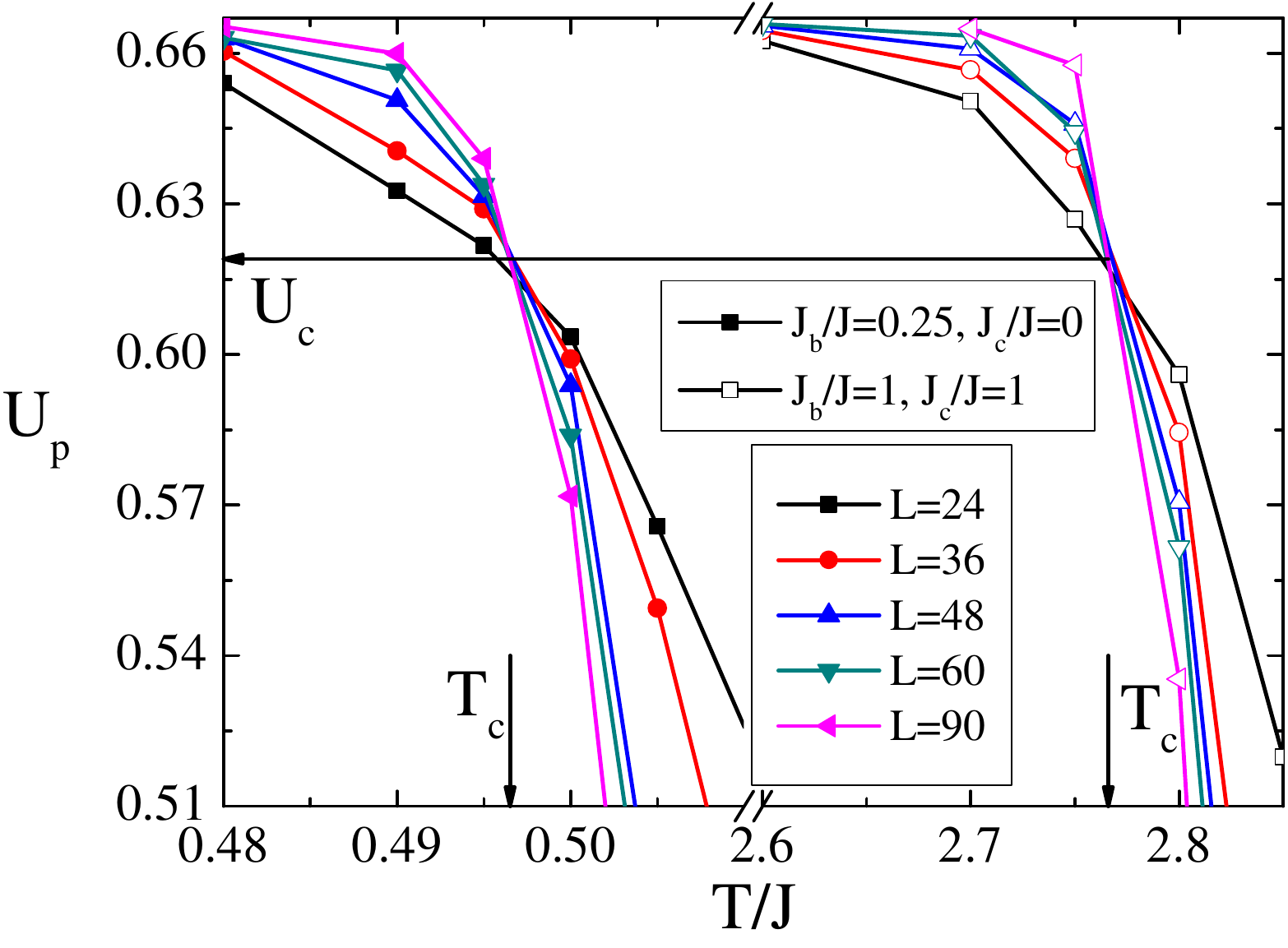}
b)
\includegraphics[scale=0.34]{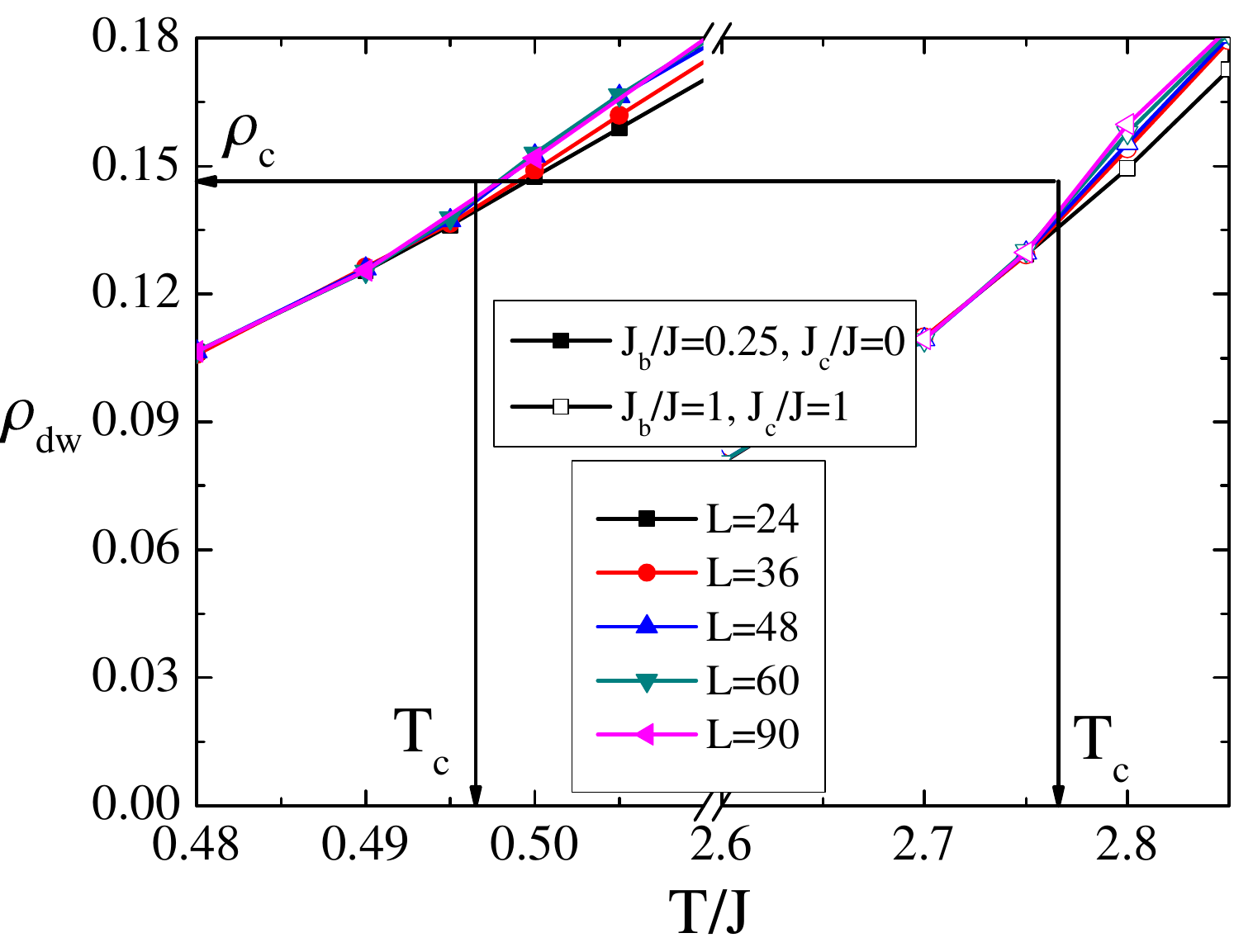}
\caption{\label{fig10} Binder cumulants (a) and wall density (b) in the Ising-XY model on a square lattice.}
\end{figure}%
In the case $J_c=0$, $J_b=0.25$, we obtain the following values of the transitions temperatures
\begin{equation}
    T_\mathrm{dw}(\square)=0.4965(5),\quad T_\mathrm{dw}(\triangle)=0.792(1),
\end{equation}
\begin{equation}
    T_\mathrm{v}(\square)=0.907(2),\quad T_\mathrm{v}(\triangle)=1.435(8),
\end{equation}
and the critical values of the topological defect densities are
\begin{equation}
    \rho_\mathrm{dw}(\square)=0.1440(30),\quad \rho_\mathrm{xj}(\square)=0.0050(10),\quad \rho_\mathrm{dw}(\triangle)=0.164(3),
\end{equation}
\begin{equation}
    \rho_\mathrm{v}(\square)=0.0087(3),\quad \rho_\mathrm{v}(\triangle)=0.0085(3).
\end{equation}
In the case $J_c=J_b=1$, we find
\begin{equation}
    T_\mathrm{dw}(\square)=2.770(8),\quad T_\mathrm{dw}(\triangle)=4.36(1),
\end{equation}
\begin{equation}
    T_\mathrm{v}(\square)=1.763(8),\quad T_\mathrm{v}(\triangle)=2.80(2),
\end{equation}
\begin{equation}
    \rho_\mathrm{dw}(\square)=0.1435(30),\quad \rho_\mathrm{xj}(\square)=0.0050(10),\quad \rho_\mathrm{dw}(\triangle)=0.164(3),
\end{equation}
\begin{equation}
    \rho_\mathrm{v}(\square)=0.0086(3),\quad \rho_\mathrm{v}(\triangle)=0.0084(3).
\end{equation}
The both cases results are in good agreement with the results for the pure Ising and $O(2)$ models.
	
Actually, we don't expect that this agreement remains good when the transitions become not well-separated in temperature. More precisely, we admit the possibility that the critical vortex density at the BKT transition differs from the universal value if the transitions are not well-separated. If $J_c\approx1$, $T_\mathrm{v}\ll T_\mathrm{dw}$, the domain wall density is too small at the BKT transition point and does not affect the BKT behavior and vortex properties. If $J_b\ll J$, discrete $\sigma$ and continuous $\bfs$ spins are almost decoupled and do not change the critical behavior in both transitions. (The case $J_b\gg J$ is dual to the case $J_b\ll J$. Since $\sigma^2=1$, the duality transformations is the simultaneous replacement $J_b\leftrightarrow J$ and $\bfs\leftrightarrow\bfk=\sigma\bfs$.) But at first sight, it is not obvious that in the case $J_c\approx0$ and $J_b\approx J$ the critical properties of domain walls and vortices remain universal.

\begin{figure}[t]
\center
a)
\includegraphics[scale=0.34]{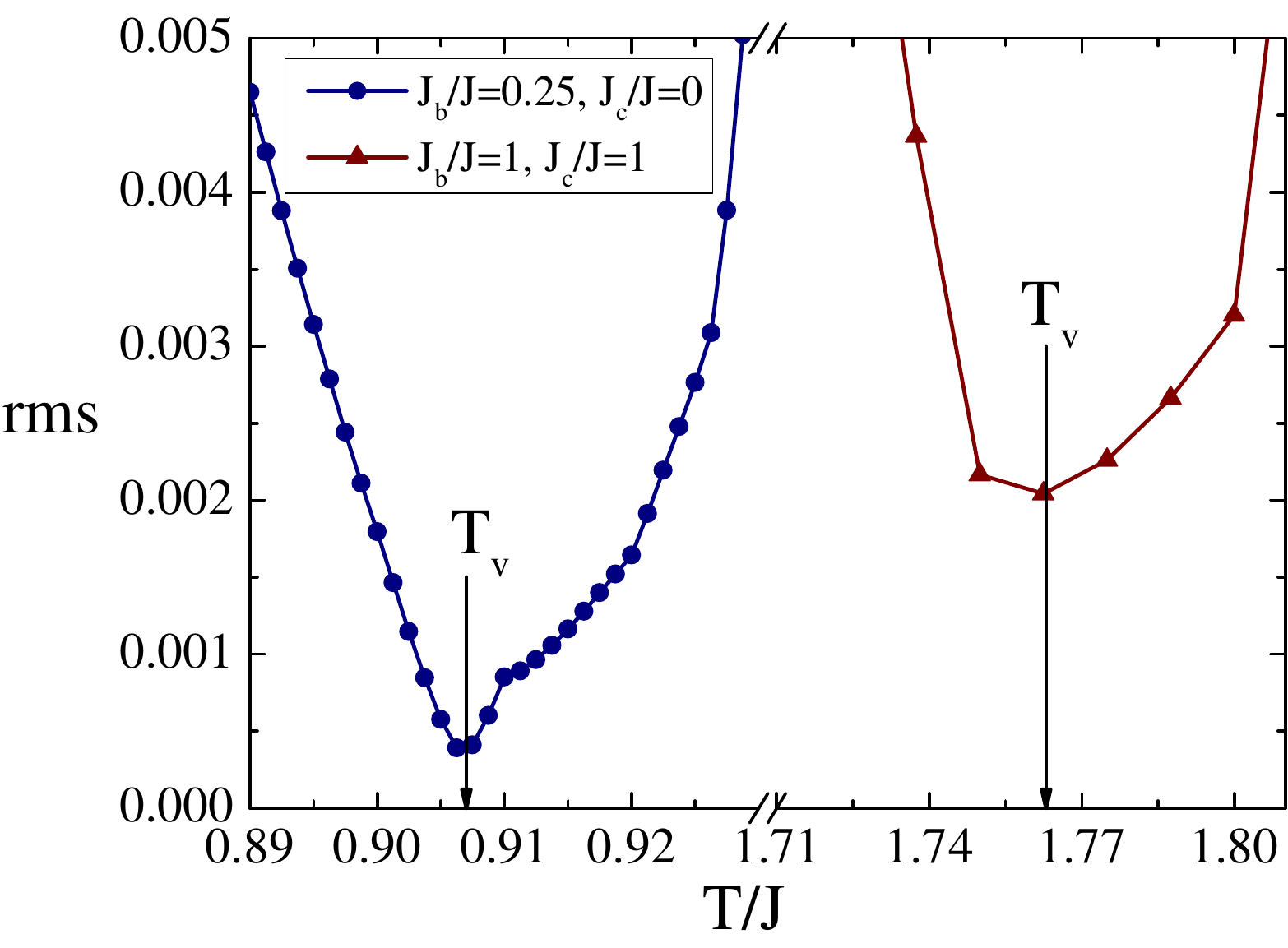}
b)
\includegraphics[scale=0.34]{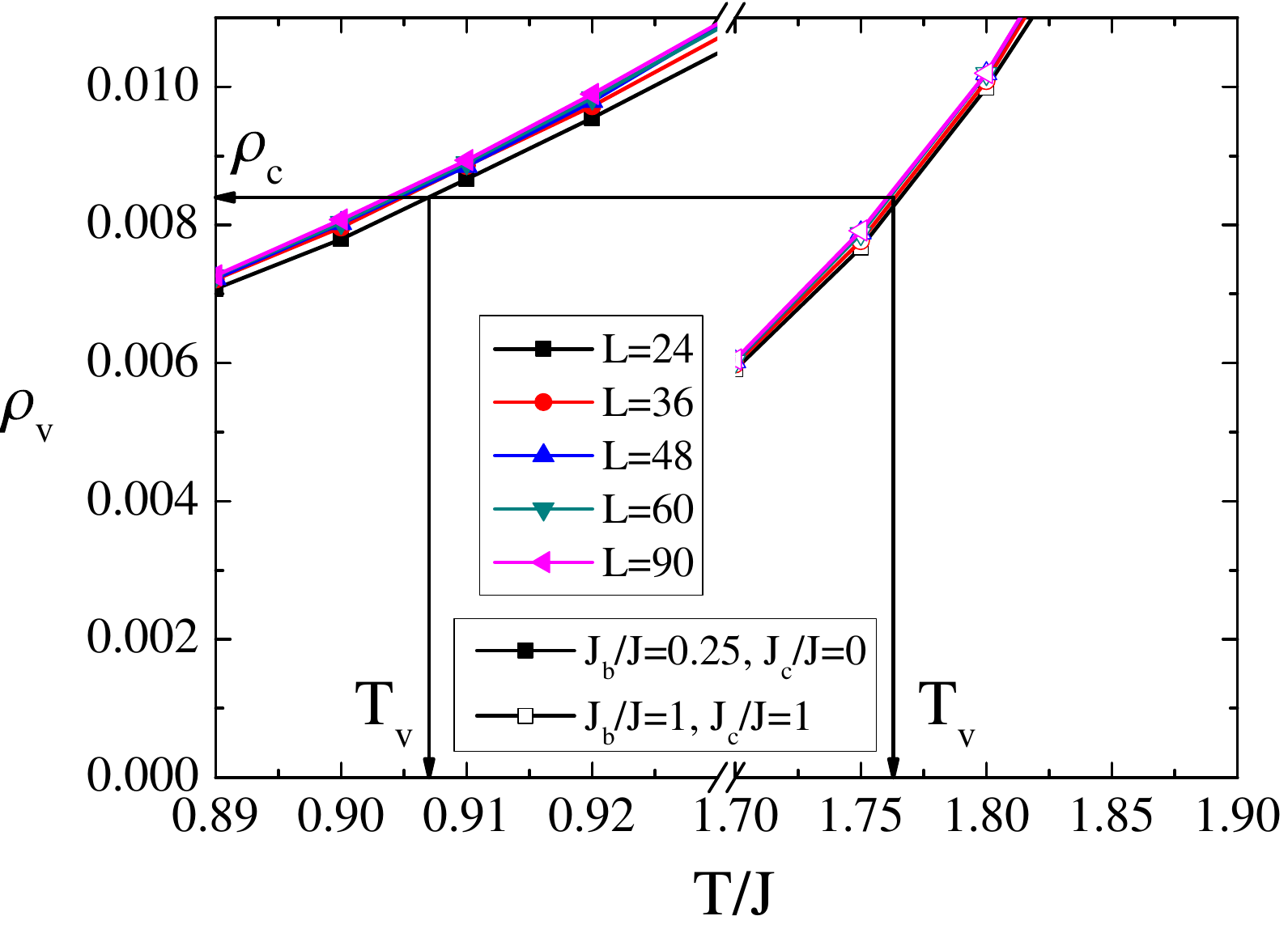}
\caption{\label{fig10} Root-mean-square error of the helicity modulus fit near the BKT transition (a) and vortex density (b) in the Ising-XY model on a square lattice.}
\end{figure}%
There is one more reason for such doubts. At $J_c\approx0$ and $J_b\approx J$, the Ising-XY model describes the critical phenomena in such systems as a Josephson junctions array in a perpendicular magnetic field, triangular XY antiferromagnet \cite{Granato87,Granato91,Kosterlitz91} or XY frustrated helimagnet \cite{Sorokin12}. An important feature of these and similar systems is the presence in the topological excitation spectrum of fractional vortices, which are corners or kinks of domain walls (depending on a specific model) \cite{Halsey85,Korshunov86,Korshunov86-2,Villain91,Uimin94}. The effective logarithmical interaction of kinks is weaker than the interaction of conventional vortices and leads to a BKT-like phase transition on a domain wall at $T_\mathrm{fv}<T_\mathrm{v}$. At $T>T_\mathrm{fv}$, a domain wall turns opaque for the correlations of a phase parameter $\varphi$ describing spin orientation $\bfs=(\cos\varphi,\sin\varphi)$. (In the case $J_b=J$ of the Ising-XY model, the domain wall opacity arises naturally, without some fractional vortices and a phase transition on a wall.) As a consequence, on approaching a Ising-like transition, the quasi-long-range order has to break down, and a BKT transition has to occurs at $T_\mathrm{v}<T_\mathrm{dw}$ \cite{Korshunov02}.

One can expect that in the presence of fractional vortices and $T_\mathrm{fv}<T_\mathrm{dw}$, the critical vortex density at a BKT transition and the critical domain wall density at an Ising transition have values larger than the universal value. For vortices, the main reason is that non-zero density of domain walls at $T>T_\mathrm{fv}$ leads to non-zero density of additional fractional vortices. For domain walls, dissociated kink-antikink pairs make a wall rugose, increasing the wall length. However, these arguments are not rigorous and may turn to be inconclusive. In particular, at $T=T_\mathrm{dw}$, the correlation between discrete $\sigma$ and continuous $\bfs$ spins as well as the wall tension vanish, so the presence of fractional vortices becomes less essential and the wall length can be any without changing the free energy. In other words, at $T_\mathrm{fv}<T<T_\mathrm{dw}$ the wall-vortex interaction is expected to change the wall dynamics as well as the wall length. But close to the critical point $T_\mathrm{dw}$, this interaction is insignificant, and the wall dynamics is determined by the critical fluctuations. Moreover, as we have noted above, the critical length of a wall (i.e. the fractional dimension in the thermodynamical limit $L\to\infty$) is uniquely determined by the conformal symmetry and the Schramm-Loewner evolution as long as the conformal symmetry remains unbroken, and a phase transition is of second order. This can serve as an important argument in favor of our hypothesis, at least for second-order transitions. Unfortunately, there are no similar arguments for vortices, and we can not exclude that the critical vortex density at a BKT transition may have a non-universal value.

The case $J_c\approx0$ and $J_b\approx J$ of the Ising-XY model will be considered in details in further studies.

\section{Ising-$V_{3,2}$ and $V_{3,3}$ Stiefel models}

By analogy with the Ising-XY model, one can easily propose a model with a domain wall - $\mathbb{Z}_2$ vortex interaction. The simplest model is the $V_{3,3}$ Stiefel model (\ref{Stiefel}). As we have shown in \cite{Sorokin17}, the transition induced by domain walls and the crossover induced by $\mathbb{Z}_2$ vortices occur at the same temperature as a first-order transition. It means that a wall-vortex interaction is non-trivial and can influence on the transition temperature as well as on the critical behavior and a type of the transition.

In this paper, we are interested in cases with well-separated transition temperatures. To investigate these cases, we introduce the Ising-$V_{3,2}$ model with the Hamiltonian
\begin{equation}
  H=-\sum_{ij} \left((J+J_b\sigma_i\sigma_j)\tr\,\Phi_i^T\Phi_j+J_c\sigma_i\sigma_j\right),\quad \Phi_i=(\bfs_i,\bfk_i), \quad \sigma_i=\pm1.
\end{equation}
In this model we consider the case $J_b=J_c=J=1$ corresponding to $T_\mathrm{dw}>T_{\mathbb{Z}_2}$. For the case with the inverse sequence of transitions $T_\mathrm{dw}<T_{\mathbb{Z}_2}$, we introduce another variant of the model:
\begin{equation}
    H=-J\sum_{ij}\left(\tr\,\Phi_i^T\Phi_j+J_b\sigma_i\sigma_j\bft_i\bft_j\right), \quad \bft_i=\bfs\times\bfk.
    \label{Stiefel}
\end{equation}
This model is an interpolation between the $V_{3,2}$ ($J_b=0$) and $V_{3,3}$ ($J_b=1$) Stiefel models. We call this model here the $V_{3,3}$ model too. We consider the case $J=1$ and $J_b=0.5$ corresponding to $T_\mathrm{dw}<T_{\mathbb{Z}_2}$. This sequence of transitions is realized in the original Ising-$V_{3,2}$ model only when $J_b\ll J$ and an wall-vortex interaction is too weak.

The study of the Ising-$V_{3,2}$ and $V_{3,3}$ models has been performed analogously to the study of the $V_{3,2}$ and Ising-XY models.

\begin{figure}[t]
\center
a)
\includegraphics[scale=0.34]{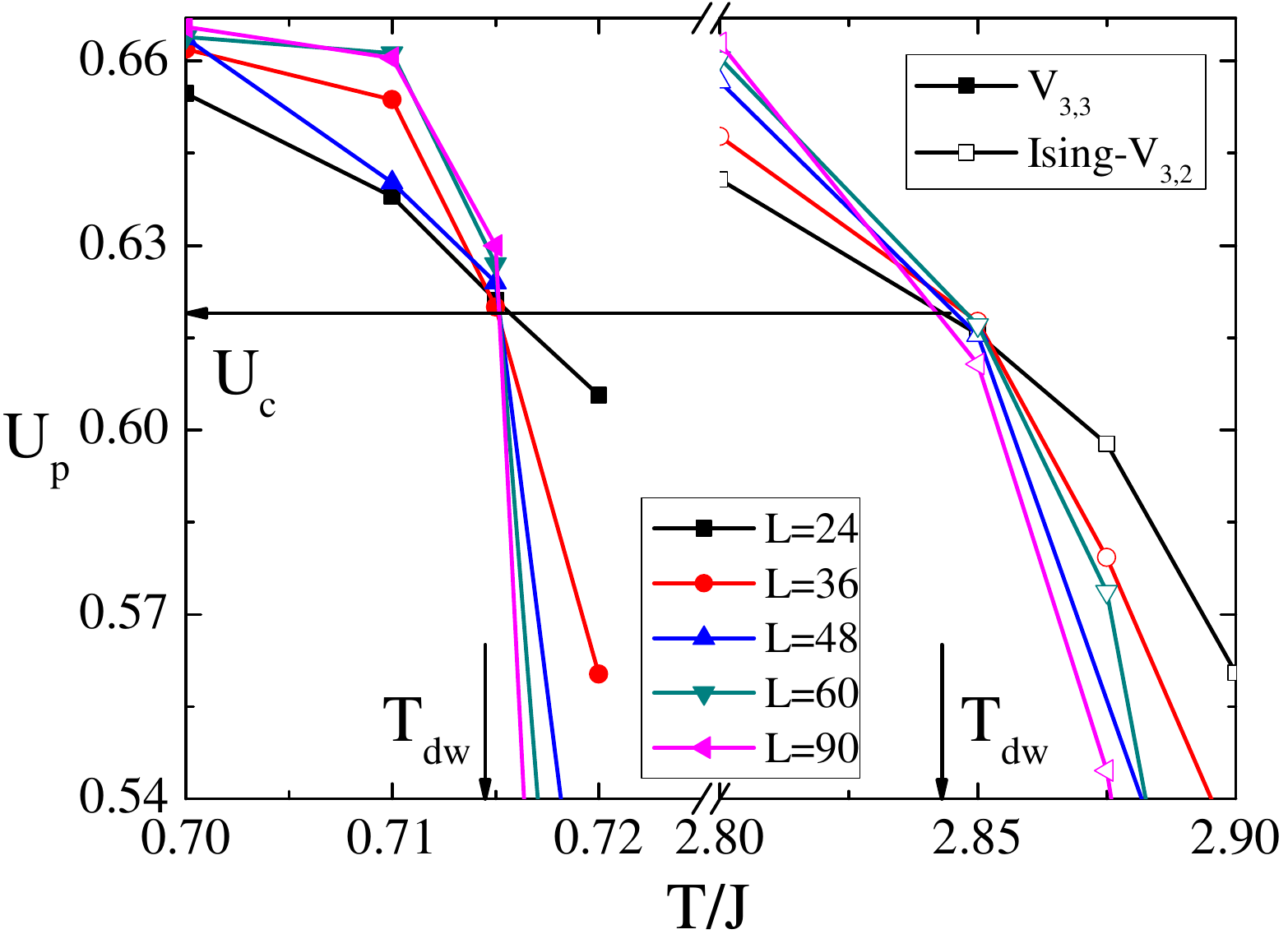}
b)
\includegraphics[scale=0.34]{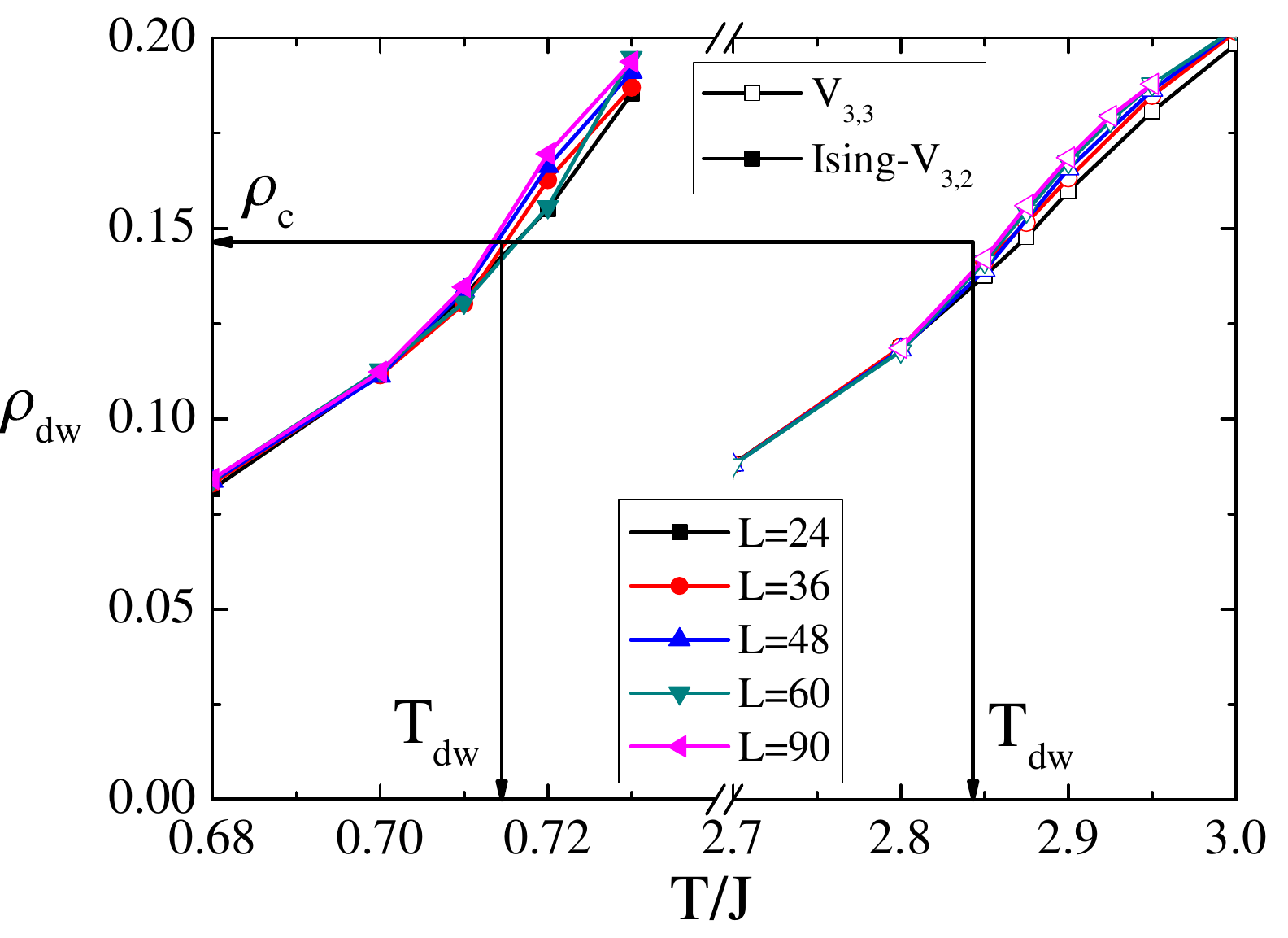}
\caption{\label{fig11} Binder cumulants (a) and wall density (b) in the $V_{3,3}$ and Ising-$V_{3,2}$ models on a square lattice.}
\end{figure}%
\begin{figure}[t]
\center
a)
\includegraphics[scale=0.34]{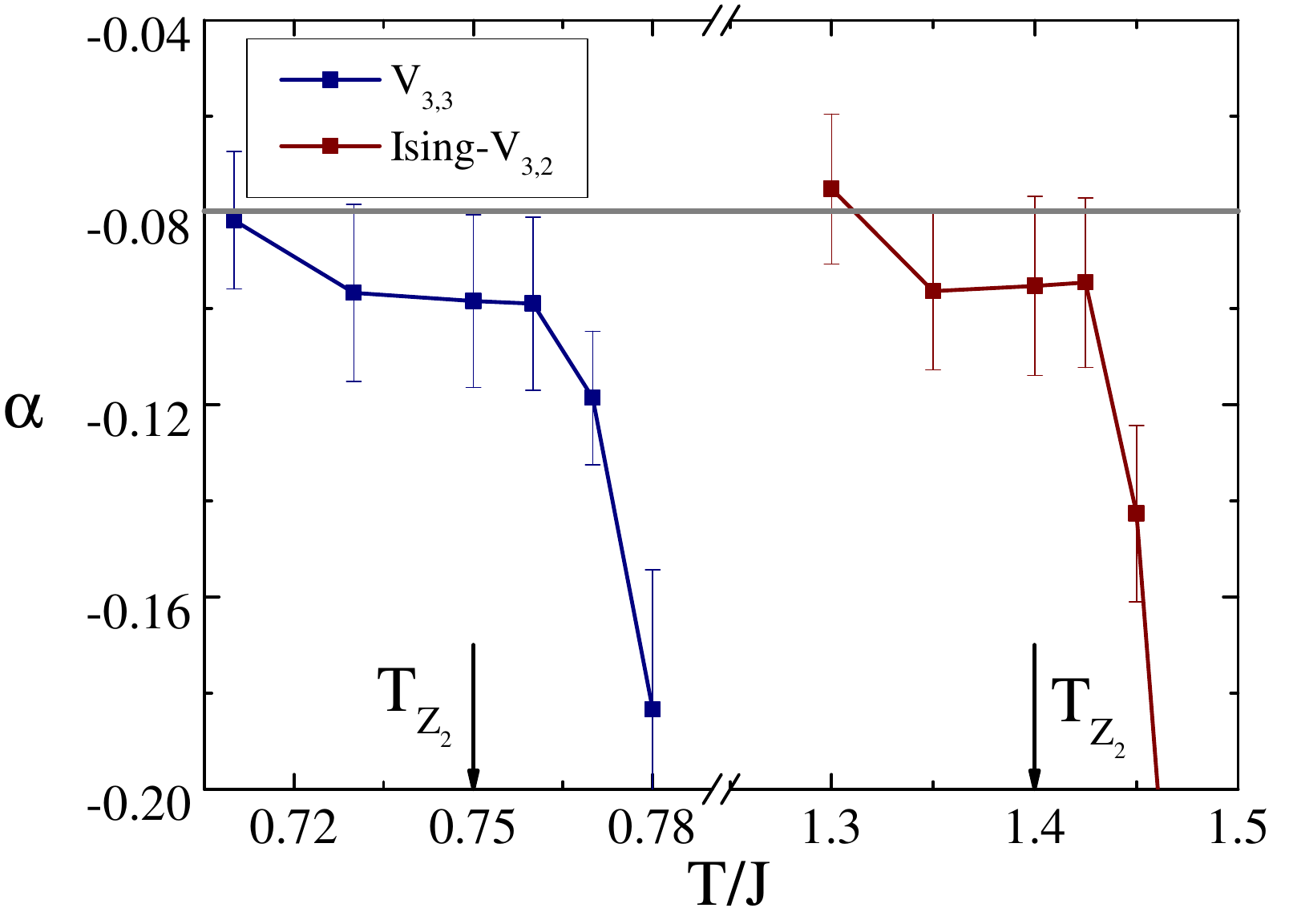}
b)
\includegraphics[scale=0.34]{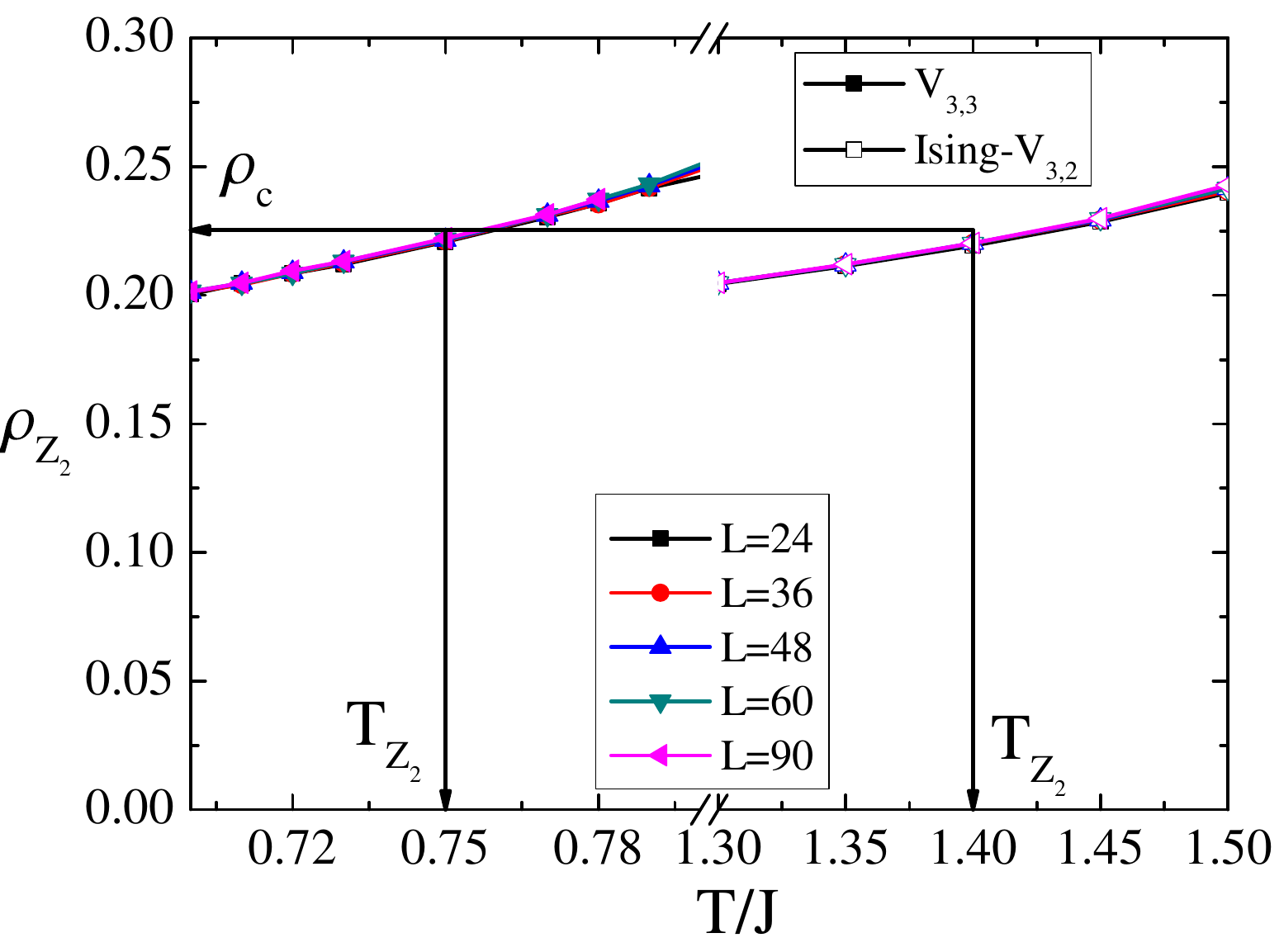}
\caption{\label{fig12} Slope of the linear fit for the helicity modulus (\ref{azaria}) near the crossover (a) and $\mathbb{Z}_2$ vortex density (b) in the $V_{3,3}$ and Ising-$V_{3,2}$ models on a square lattice.}
\end{figure}%
For the Ising-$V_{3,2}$, we find (see figs. \ref{fig11} and \ref{fig12})
\begin{equation}
    T_\mathrm{dw}(\square)=2.843(7),\quad T_\mathrm{dw}(\triangle)=4.41(1),
\end{equation}
\begin{equation}
    T_{\mathbb{Z}_2}(\square)=1.40(3),\quad T_{\mathbb{Z}_2}(\triangle)=2.25(4).
\end{equation}
The critical values of the topological defect densities are
\begin{equation}
    \rho_\mathrm{dw}(\square)=0.1450(30),\quad \rho_\mathrm{xj}(\square)=0.0048(10),\quad \rho_\mathrm{dw}(\triangle)=0.164(4),
\end{equation}
\begin{equation}
    \rho_{\mathbb{Z}_2}(\square)=0.223(8),\quad \rho_{\mathbb{Z}_2}(\triangle)=0.234(8).
\end{equation}
For the $V_{3,3}$ model,
\begin{equation}
    T_\mathrm{dw}(\square)=0.7145(5),\quad T_\mathrm{dw}(\triangle)=1.304(8),
\end{equation}
\begin{equation}
    T_{\mathbb{Z}_2}(\square)=0.750(20),\quad T_{\mathbb{Z}_2}\triangle)=1.31(20),
\end{equation}
\begin{equation}
    \rho_\mathrm{dw}(\square)=0.1490(30),\quad \rho_\mathrm{xj}(\square)=0.0044(10),\quad \rho_\mathrm{dw}(\triangle)=0.170(4),
\end{equation}
\begin{equation}
    \rho_{\mathbb{Z}_2}(\square)=0.225(3),\quad \rho_{\mathbb{Z}_2}(\triangle)=0.236(8).
\end{equation}
All results coincide within the margin of error with the results for the pure Ising and $V_{3,2}$ models.

Note that the case $J_c\approx0$ and $J_b\approx J$ in the Ising-$V_{3,2}$ model has similar properties as it in the Ising-XY model, namely that a domain wall induces disorder in the continuous order parameter. This case will be also considered in further studies.

\section{Vortex-vortex interactions}

To investigate a possible vortex-vortex interaction for the both types of vortices, one can consider the XY-XY, XY-$V_{3,2}$ and $V_{3,2}$-$V_{3,2}$ models formulated in a spirit of the Ising-XY and Ising-$V_{3,2}$ models. In particular, the XY-XY model is
\begin{equation}
  H=-\sum_{ij} \left((J+J_b\bfk_i\bfk_j)\bfs_{i}\bfs_{j}+J_c\bfk_i\bfk_j\right),
\end{equation}
where $\bfs_{i}=(\cos\varphi_i,\sin\varphi_i)$ and $\bfk_{i}=(\cos\psi_i,\sin\psi_i)$. But in contrast to the Ising-XY model, one order parameter does nor perceive a topology of another order parameter. Thus, such a vortex-vortex interaction turns to be trivial. It changes the transition temperatures but does not change the critical behavior. In particular, a possible multicritical point has the critical behavior equivalent to the case of two decoupled XY models. The XY-$V_{3,2}$ and $V_{3,2}$-$V_{3,2}$ models are trivial too.

We can add to the XY-XY model the term $J_d(\bfs_i\bfk_i)^2$, then phases of spins $\bfs$ and $\bfk$ become concerted, and a vortex-vortex interaction is expected to be non-trivial. But such a interaction changes the symmetry of the Hamiltonian, and the XY-XY model falls into the same universality class as the Ising-XY model \cite{Kosterlitz86}.

We know the non-trivial model where the order parameter space is wanted $G/H=SO(2)\otimes SO(2)$. As we have shown in \cite{Sorokin14}, such an order parameter space is realized in a two-dimensional classic ferromagnet with the Dzyaloshinskii-Moria interaction. Unfortunately, one $SO(2)$ subgroup corresponds to rotations in the coordinate space, so in the lattice version of the model this symmetry is broken to a discrete subgroup determined by a lattice. However, we don't exclude existence other non-trivial models with two flavors of vortices and their interaction.

\section{Conclusion}

The results obtained in this paper confidently support the hypothesis that the density of topological defects has an universal critical value at a corresponding continuous transition point. All considered cases demonstrate independence of the critical density value on a lattice type (with the obvious corrective for a unit length of line-like defects discussed in the section devoted to the Ising model). More importantly that the hypothesis is confirmed not only for simplest pure models with a single transition and one type of defects, but also for models with a few types of topological excitations and successive transitions. For clarity, we place all the obtained values of defect densities in tables \ref{tab1}, \ref{tab2} and \ref{tab3}.
\begin{table}[t]
\center
\small
\caption{Comparison of domain wall density at Ising-like transition points of different models. The density ratio $r_\mathrm{dw}$ comparing results for a square and triangular lattices is defined as (\ref{ratio-ising-new}).}
\label{tab1}
\begin{tabular}{cc|cccc}
\hline
\hline
Model & Class $G/H$ &   $\rho_\mathrm{dw}(\triangle)$ & $\rho_\mathrm{dw}(\square)$& $\rho_\mathrm{xj}$ & $r_\mathrm{dw}$\\
\hline
$V_{1,1}\equiv$Ising & $\mathbb{Z}_2$ &  0.16663(3) & 0.14644(2) & 0.00494(2) & 1.0001(2) \\
AT, $\langle s\rangle$   & $\mathbb{Z}_2\otimes\mathbb{Z}_2$ &  0.167(1) & 0.1470(7) & 0.0055(7) & 1.000(10) \\
AT, $\langle s\sigma\rangle$& $\mathbb{Z}_2\otimes\mathbb{Z}_2$ & 0.166(1) & 0.1461(8) & 0.0050(8) & 1.001(10) \\
$V_{2,2}$, $T_\mathrm{v}>T_\mathrm{dw}$& $\mathbb{Z}_2\otimes SO(2)$ & 0.164(3) & 0.1440(30) & 0.0050(10) & 0.999(20) \\
$V_{2,2}$, $T_\mathrm{v}<T_\mathrm{dw}$& $\mathbb{Z}_2\otimes SO(2)$ & 0.164(3) & 0.1435(30) & 0.0049(10) & 0.996(20) \\
$V_{3,3}$, $T_{\mathbb{Z}_2}>T_\mathrm{dw}$& $\mathbb{Z}_2\otimes SO(3)$ & 0.170(4) & 0.1490(30) & 0.0044(10) & 0.999(20) \\
$V_{3,3}$, $T_{\mathbb{Z}_2}<T_\mathrm{dw}$& $\mathbb{Z}_2\otimes SO(3)$ & 0.164(3) & 0.1450(30) & 0.0048(10) & 1.007(20) \\
\hline
\hline
\end{tabular}
\end{table}
\begin{table}[t]
\center
\small
\caption{Comparison of vortex density at BKT-like transition points of different models.}
\label{tab2}
\begin{tabular}{cc|ccc}
\hline
\hline
Model & Class $G/H$ &  $\rho_\mathrm{v}(\triangle)$ & $\rho_\mathrm{v}(\square)$ & $r_\mathrm{v}$ \\
\hline
$V_{2,1}\equiv O(2)$ & $SO(2)$ &  0.0083(3) & 0.0085(3) & 1.02(3) \\
$V_{2,2}$, $T_\mathrm{v}>T_\mathrm{dw}$& $\mathbb{Z}_2\otimes SO(2)$ & 0.0085(3) & 0.0087(3) & 1.02(3)  \\
$V_{2,2}$, $T_\mathrm{v}<T_\mathrm{dw}$& $\mathbb{Z}_2\otimes SO(2)$ & 0.0084(3) & 0.0086(30) & 1.02(3)  \\
\hline
\hline
\end{tabular}
\end{table}
\begin{table}[t]
\center
\small
\caption{Comparison of $\mathbb{Z}_2$ vortex density at crossover points of different models.}
\label{tab3}
\begin{tabular}{cc|ccc}
\hline
\hline
Model & Class $G/H$ &  $\rho_{\mathbb{Z}_2}(\triangle)$ & $\rho_{\mathbb{Z}_2}(\square)$ & $r_{\mathbb{Z}_2}$ \\
\hline
$V_{3,2}$ & $SO(3)$ &  0.229(15) & 0.221(10) & 0.97(6) \\
$V_{3,3}$, $T_{\mathbb{Z}_2}>T_\mathrm{dw}$& $\mathbb{Z}_2\otimes SO(3)$ & 0.236(15) & 0.225(10)  & 0.95(6) \\
$V_{3,3}$, $T_{\mathbb{Z}_2}<T_\mathrm{dw}$& $\mathbb{Z}_2\otimes SO(3)$ & 0.234(15) & 0.223(10)  & 0.95(6) \\
\hline
\hline
\end{tabular}
\end{table}

Of cause, the considered models do not cover all possibilities, but we consider perhaps the most representative cases including a second-order phase transition, a BKT-like transition and even an expressed crossover not being even a true transition. There are a number of cases among not considered ones useful as hypothesis tests because they potentially contain difficulties refuting the hypothesis. One of such cases we have mentioned above in context of the Ising-XY model when topological excitations of two types interact non-perturbatively and lead to appearance of an additional type of defects. Another example also noted above is the 3- and 4-state Potts models, where X- and Y-junctions of domain walls are expected to be relevant to a transition and can be considered as new defects.

One can easily list more of potentially difficult cases. First, site-diluted models contains domain-wall-like bounds separating clusters of magnetic and non-magnetic sites. It is not obvious, e.g., how these bounds affect the critical properties of domain walls in the site-diluted Ising model. (An example of difficulties has been discussed in \cite{Najafi16}, see however \cite{Sorokin17-2}.) A similar situation arises if the local definition of an order parameter is not non-normalized and admits a zero value. Thus, a non-perturbed domain wall in a frustrated XY helimagnet passes through lattice bonds with a zero chirality and changes its position under an arbitrarily small perturbation. It seems like a wall has an additional "soft mode"\ disappearing under thermal fluctuations. Of cause, this "internal mode"\ is artefact of the definition of a chiral order parameter and a wall position, but it inevitably arises in models with a non-normalized order parameter. Such a situation is typical for models of frustrated magnets.

The last example should be extended to the more general case when a topological defect has an extra internal degrees of freedom or non-zero core size. The most fundamental example in this case is a spontaneously breaking of a {\it gauge} symmetry. If a broken symmetry $G/H$ is local, a core of corresponding topological defect has non-vanishing size proportional to a gauge coupling constant. However, we assume that the case of a gauge symmetry does not always lead to a refutation of the hypothesis.

It is generally accepted that a phase transition with a gauge symmetry breaking falls to the same universality class as a transition with a corresponding global symmetry (see however interesting recent work \cite{Pelissetto17}). It is because the critical behavior is described by large-scale fluctuations with a weak dependence on spatial coordinates, so a local symmetry can be considered as a global one. Another way to reduce local symmetry to global one is to deal with the limit of a vanishing gauge coupling. In particular, this limit can be taken as a good approximation if the Higgs coupling constant $\lambda$ responsible for a symmetry breaking is much larger than a gauge coupling $g$, $\lambda\gg g$. Such a situation takes place in type-II superconductors. But when a gauge coupling constant vanishes, a size of a defect core vanishes too, so in the limit $g\to0$ we restore the usual picture of a model with a global symmetry.

However if a gauge coupling is large enough, an effective defect-defect interaction may changes its form crucially. For example, exotic configurations with a large topological charge may become more energetically favorable than usual single-charge defects. Remaining relevant to a phase transition, these novel topological excitations may influence the critical behavior as a non-perturbative effect. A least, the density of single-charge defects must be non-universal.

Finally, we have excluded skyrmion-like defects for our consideration. Defects of this type are dissimilar to defects of other types. They have no cores where a full symmetry $G$ is unbroken, and we can not claim that a large concentration of skyrmions leads to an order destruction. On the other hand, we can not exclude that skyrmions are relevant to a phase transition and may change the critical behavior.

But even taking into account the mentioned difficulties, our hypothesis is extremely useful, since it can be used to determine a transition temperature as well as a type of critical behavior and universality class.
\medskip

This work is supported by the RFBR grant No 16-32-60143.

\end{document}